\documentclass[preprint,aps,showpacs,floats,superscriptaddress,nofootinbib]{revtex4}
\usepackage{amsmath}
\usepackage{amssymb}
\usepackage{graphicx}
\newcommand{\Mpl}{M_\mathrm{pl}}
\newcommand{\gravitino}{\tilde{G}}
\newcommand{\mgravitino}{m_{\tilde{G}}}
\newcommand{\axino}{\tilde{a}}
\newcommand{\maxino}{m_{\tilde{a}}}
\newcommand{\lsim}{\buildrel<\over{_\sim}}
\newcommand{\gsim}{\buildrel>\over{_\sim}}
\newcommand{\mgl}{m_{\tilde{g}}}
\newcommand{\mchii}{m_{\tilde{\chi}^0_1}}
\newcommand{\mstau}{m_{\tilde{\tau}}}
\newcommand{\stau}{\tilde{\tau}}
\newcommand{\xg}{x_{\gamma}}
\newcommand{\ct}{\cos\theta}
\newcommand{\ab}{A_{\tilde{B}}}
\newcommand{\bunbo}{
x_{\gamma}(1+\cos\theta)-2A_{\tilde{B}}+A_{\tilde{B}}x_{\gamma}(1-\cos\theta)
}

\begin{document}
\preprint{UT-06-23}
\preprint{KEK-TH-1119}
\title{Prospects to study a long-lived charged next lightest 
supersymmetric 
particle at the LHC
}
\author{Koichi Hamaguchi} 
\affiliation{Department of Physics, University of Tokyo, Tokyo 113-0033, Japan}
\author{Mihoko M. Nojiri}
\affiliation{IPNS, KEK, Oho 1-1, 305-0801, Japan}
\author{Albert de Roeck}
\affiliation{Physics Department, CERN, CH-1211 Geneva 23, Switzerland}
\affiliation{Universitaire Instelling Antwerpen,
B-2610 Wilrijk, Belgium}


\begin{abstract}
If the scalar tau $\stau$ is the next lightest supersymmetric 
particle and  decays into a gravitino($\gravitino$) 
being the lightest supersymmetric particle, it will have 
generally a very long lifetime.
In this paper, we investigate the possibility to study the
decay of such a  long lived scalar tau at the LHC.   If we
can add to the present LHC experiments additional detectors which are able 
to stop the stau particles and measure
the produced decay products,  the decay 
characteristics can be studied precisely at the LHC.
We identify a maximum "stopper detector" that could  be added in
the CMS cavern, and estimate the sensitivity to the
lifetime of the stau and to the mass of gravitino with this detector.
The decay of the scalar tau may be significantly modified if the decay
channel to the axino $\axino$ is open.
We study the possibility to distinguish such decays from
decays into gravitinos by measuring the process
$\stau\rightarrow \axino (\gravitino) \tau \gamma$  using the stopper
detector.
\end{abstract}


\maketitle

\section{introduction}
The minimal supersymmetric standard model (MSSM) is one of the most
important candidates for the physics beyond the Standard Model. The
strongly interacting superpartners, gluinos and squarks with mass
lighter than 2.5 TeV can be discovered at the Large Hadron Collider at
CERN (LHC)~\cite{ATLASTDR}.  The  physics run of LHC will 
start in 2008. 

Among the supersymmetric (SUSY) particles, the lightest SUSY particle
(LSP) plays a key role. In cosmology, it is a candidate for the cold
dark matter in the universe. At the LHC, the signals of the
supersymmetric particles depends on the nature of the LSP.  It may be
the lightest neutralino $\tilde{\chi}^0_1$, which escapes from
detection and leads to missing $E_T$ in the event.  Another
possibility for the LSP is the gravitino $\gravitino$, 
the superpartner of the
graviton.  The gravitino coupling to  other particles in the MSSM
sector is extremely small, namely proportional to $1/\Mpl$.  The
next lightest SUSY particle (NLSP), whose decay products necessarily
include the gravitino, is therefore long-lived.

We discuss the scenario where the LSP is the gravitino and the
long-lived NLSP carries charge.  A natural
candidate for a charged NLSP (CNLSP) is the lightest scalar tau,
$\tilde{\tau}_1$, which can be 
significantly lighter than the other sleptons due to the left-right
mixing induced by the off diagonal matrix element $m^2_{LR}\sim
m_{\tau}\mu\tan\beta$.  The charged particle leaves tracks in  the
central detectors (ATLAS and CMS), giving additional information for the
SUSY particle reconstructions.  If the NLSP decays in the main detectors, 
a displaced vertex may be observed as well. 

The expected lifetime of the CNLSP $\tilde{\tau}$ is unconstrained,
because it is proportional to $(\mgravitino)^2$
 the yet unknown gravitino mass squared.  On the other hand,
the gravitino mass is proportional to the total SUSY breaking scale in
the hidden sector, therefore the determination of the lifetime is 
 an important physics goal.  The lifetime measurement gives 
direct information on the hidden sector.

A particle decays efficiently in the main detectors (CMS and ATLAS) if
the decay length is sufficiently short, $c\tau\ll$ (10m)$ \times
N_{\rm produced}$, where $N_{\rm produced}$ is the number of produced
SUSY particles.  On the other hand, for typical SUSY production cross 
sections,
a direct observation of the decay is very difficult for
$\tau_\mathrm{CNLSP}> 0.01$ sec.
However, it has been pointed out  that the CNLSP stopped by the
ionization loss in the material may be studied in
detail~\cite{Hamaguchi:2004df,Feng:2004yi}. An idea for a stopper 
based on a water tank 
is presented in Ref.\cite{Feng:2004yi}, where the 
water can be
transported away from the detector site, for concentration and further
study.  In Ref.~\cite{Hamaguchi:2004df}, a detector consisting of a 
tracker and heavy stopping material is proposed, which can measure the
arrival time and the location where the CNLSPs are stopped, in 
addition to the energy of the decay products.
Another possibility, which requires minimal experimental modification, is
to study the CNLSP which are stopped in the main detector or surrounding
rock~\cite{albert}.

As  pointed out in Ref.~\cite{Buchmuller:2004rq}, the 
study of the CNLSP decays can probe the underlying
supergravity in nature. With the gravitino mass inferred from the 
kinematics, the additional CNLSP lifetime measurement will test an 
unequivocal
prediction of supergravity. Moreover, the study
of a rare 3--body decay $\stau\to\tau\gamma\gravitino$ can  reveal
the peculiar couplings of the gravitino and the
gravitino spin $3/2$.

In this paper, we consider the physics of the CNLSP $\tilde{\tau}$ decays 
that can be done with the heavy
material stopper--detector~\cite{Hamaguchi:2004df}, because only a
detector of this type can cover a wide range of lifetime $O(10{\rm nsec}) $
$<\tau_\mathrm{CNLSP}< \mathcal{O}(10$ years).
We find that
the mass of the gravitino 
can be
measured 
if it is sufficiently heavy (roughly $m_{\tilde{G}}>0.2 m_{\tilde{\tau}}$).  
In that case, one can check 
if the lifetime is consistent with the supergravity interpretation.

As we shall see, the LSP mass resolution is however poor if 
$m_{\tilde{G}} \lsim 0.2\mstau$ .
In that case, it is very hard to prove that the decay  $\tilde{\tau}
\rightarrow \tau$ plus an  invisible particle $X$  is indeed 
caused by the supergravity interaction involving $\tilde{G}$.
We should note that, because of the extremely weak coupling of
the gravitino, if there is any unknown supersymmetric particle $X$
which couples rather weakly to the MSSM particles, the lightest MSSM
particle may decay into $X$ instead of $\gravitino$ even if the
$\gravitino$ is the LSP.  For example, the superpartner of the axion, the
axino $\axino$, can be such a particle. 
 However, we found that it may be possible to discriminate the
case of $X = \gravitino$ and the case of $X = \axino$, by
investigating the three body decay $\tilde{\tau}\rightarrow \tau\gamma
X$ as suggested in \cite{Buchmuller:2004rq,Brandenburg:2005he}, with 
 enough statistics.

This paper is organized as follows.  In section \ref{sec:detector}, we
discuss a possible detector setup. We found an $\mathcal{O}(1)$ kton
detector (up to 8 kton) may be placed next to the CMS detector without
serious modification of the CMS experiment itself, but 
with non-negligible
modifications to the CMS cavern side walls. In 
section
\ref{sec:modelpoints}, we select several model points and estimate the
expected number of the stopped particles for $\int {\cal L} =300$
fb$^{-1}$ and 3000 fb$^{-1}$ . In Section \ref{sec:twobody}, we
discuss the measurement of the two body decay $\tilde{\tau}\rightarrow
\tau X $.  The three body decay of the CNLSP is studied in Section
\ref{sec:three-body}.  Section \ref{sec:discussion} contains the
discussion and comments.

\section{Assumptions on the stoppers}
\label{sec:detector}

\begin{table}
\begin{center}
\begin{tabular}{|c|c|c|c|}
 \hline 
           & diameter & weight of the detector& length\cr
           \hline
ATLAS&   22m  &   7Kt &  44m\cr
CMS&  15m   &  12.5Kt& 21m \cr
\hline 
\end{tabular}\label{atlascms}
\end{center}
\end{table}
In this section, we discuss the possibility to install massive stoppers
next to the LHC detectors, CMS and  ATLAS. It turns out  that the 
CMS cavern may allow for an easier installation and more room for a massive 
stopper,  compared to the ATLAS cavern. 
The parameters of the two detectors are listed in Table.~\ref{atlascms}.  
The diameter of the CMS detector is smaller, 
 therefore the massive stopper can be 
placed closer to the
interaction point  at CMS.  The weight  of the  CMS 
detector is about twice as large as that of ATLAS.  Because of the 
large weight, 
the cavern of CMS  is designed so that it can 
sustain a massive object safely, which includes a reinforced floor
to spread the detector pressure more equally. A potential 
massive stopper with a weight of around a few kton can be placed on
both sides of the CMS detector, but it will need a reorganization of the 
scaffolding and gallery paths on the cavern walls, to make room for such 
an additional detector.

The assembly and construction of the two detectors is  also 
very different.  
Most of 
the CMS detector components are assembled  
on the ground, and about 15 large detector units will be lowered 
in the cavern for final assembly works. Hence the installation is less 
integrated with the cavern, leaving relatively more freedom and 
 thus changes needed in the cavern to install massive stoppers are 
somewhat simpler~\cite{CMS_proposal}.
On the other hand ATLAS detector is assembled mostly in
its cavern. The huge magnets toroids and outside muon system fill up the 
cavern.
The cryogenic system in the ATLAS cavern is also taking
space outside the detector~\cite{ATLAS_proposal}.

\begin{figure}[t]
\begin{center}
\includegraphics[width=7cm]{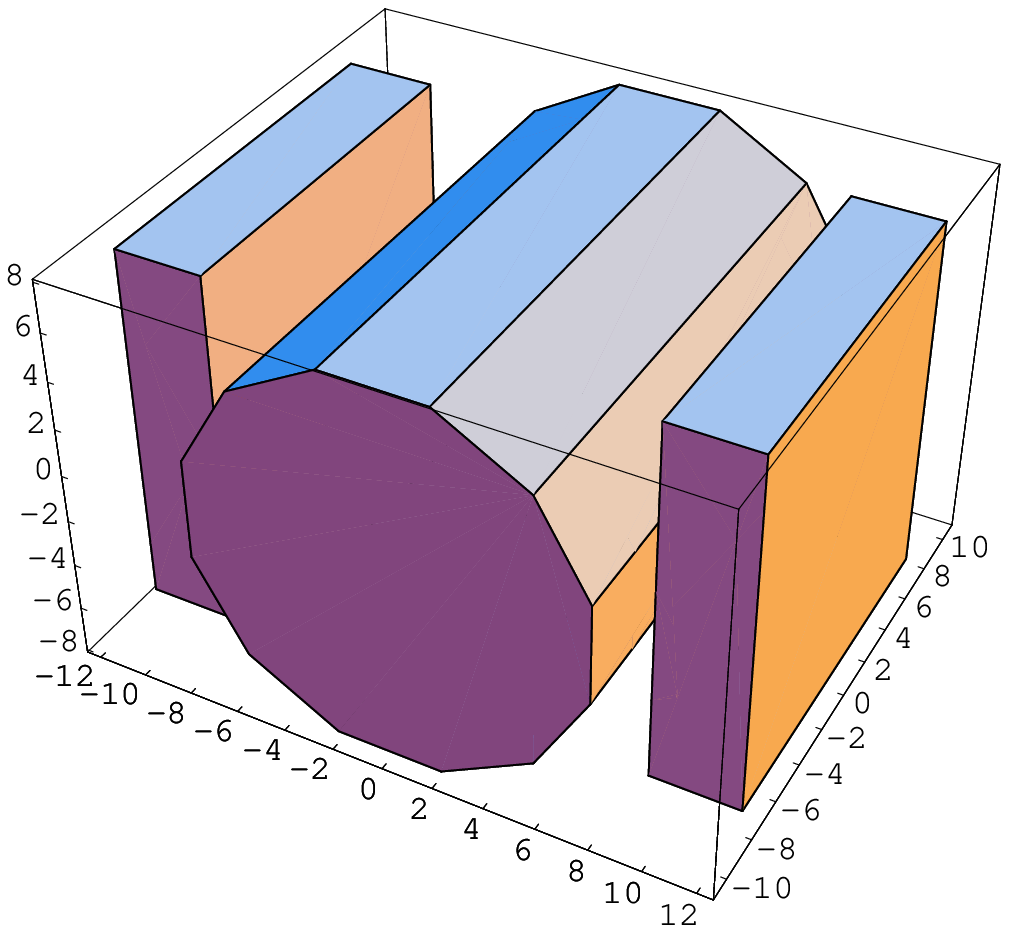}
~~~~
\includegraphics[width=7cm]{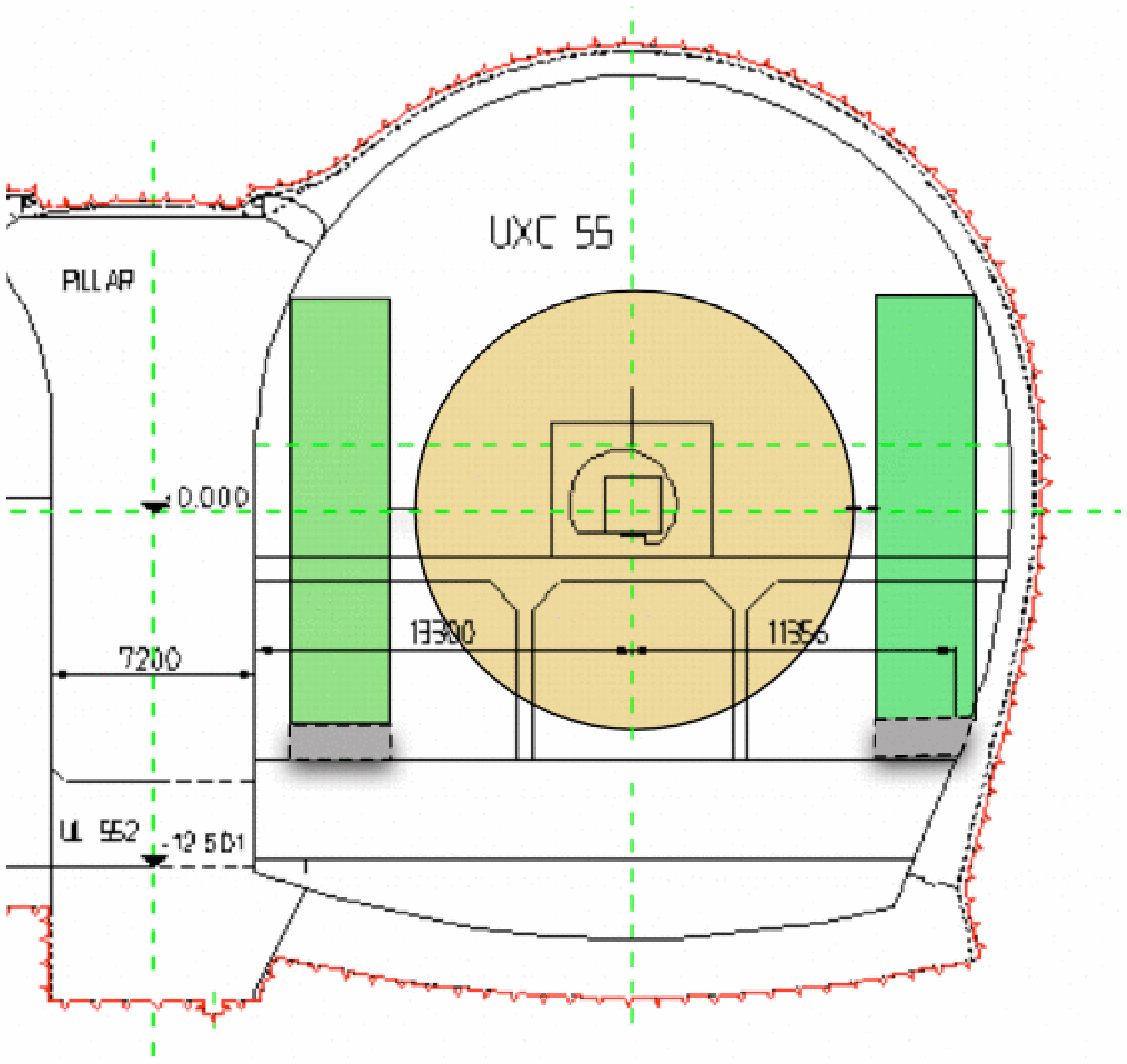}
\end{center}
\caption{
Left: a schematic figure of the CMS detector and two stoppers.
The numbers are in units of meters, and $(0,0,0)$ is  the collision point.
Right:
two stopper--detectors and a circle about the size of CMS
detector are superimposed on the cross section of CMS cavern UXC 55,
drawing taken from Ref.~\cite{CERN-STCVDC}.
}
\label{Fig:stopper}
\end{figure}

We assume two stoppers with the size 3.5m $\times$
15m $\times$ 15m and the average density $\rho_{\rm stop}= 5$g/cm$^3$, 
 hence, the total weight of the detector is 8 kton. 
This is maximum possible rectangular parallelepiped volume
that can be placed in the cavern, with the given space to the 
cavern wall and with its long edge being the same as that of the CMS barrel part
(see Fig.~\ref{Fig:stopper}).
The stoppers are thus
placed 8.5~m away from the interaction point.
We also assume that stopping power of the CMS detector is equivalent to
$(2500/\sin\theta)$ {\rm g/cm}$^2$ iron,  where $\theta$ is angle between  
the CNLSP direction and the  beam direction.  
  The number comes from the average density of CMS
detector, 3.37g/cm$^3$, which leads to the weight  per cm$^2$ for the 
radial direction of $2500{\rm g/cm}^2$.  

As discussed in the previous paper~\cite{Hamaguchi:2004df}, the stopper
can be a hadronic and electromagnetic calorimeter simultaneously, if the 
detector consists of layers of dense stopper and tracking devices. 
The measurement of the energy of the decay product of the CNLSP is the
key ingredient to explore the CNLSP interactions to the $X$ particle.
In this paper we assume that the CNLSP is the scalar tau lepton $\stau$, 
which  decays mostly as
$\stau \rightarrow \tau X$ where $X=\gravitino$ or $\axino$.

The  $\tau$ decays into $l\bar{\nu}_l \nu_{\tau} $, or into 
$\pi^\pm$ and $\pi^0$'s. 
We do not consider the decays into $\mu$, because the muon energy 
 cannot be measured 
unless the stopper contains a magnetic field.   
The energy   of the leptons are 
much softer than the parent $\tau$ energy anyway, so that they are 
less  useful for the study of the decay kinematics. 

A large volume detector is advantageous  to measure the 
energy of the $\tau$ decay products, because the 
detector must contain  most of  the energy of the showers 
from   the  $\tau$ decay products. 
To fully absorb the hadronic cascade one needs sufficient thickness
of the calorimeter.  The 
required
thickness for an iron calorimeter (density is 7.87g/cm$^3$)  is listed in
Ref.~\cite{Eidelman:2004wy,Bock:1980rs}, and is about 170 (120) cm for
100~GeV single hadron energy for 99\% (95\%) containment
respectively, equivalent to  1337.9g/cm$^2$ (944.4g/cm$^2$).  To measure
the energy with sufficient accuracy 
by a stopper with $\rho=$
5g/cm$^3$, the distance between the decay position and the end of the 
detector along the shower path  must be at least 190cm.
A simplified Monte Carlo simulation shows that  63\% of the 
$\tilde{\tau}$ decays satisfy this condition. 
High energy photons and electrons initiate electromagnetic cascades, 
which are much easier to contain in a detector. 
The energy deposition along the axis of the cascade is
well described by a gamma distribution, where the maximum occurs at $\sim
8X_0$, where $X_0$ is the radiation length.  The
energy deposition terminates around
$15X_0$. (See Fig.27.17 of \cite{Eidelman:2004wy}.)
For the case of iron ($X_0= 13.84{\rm g/ cm}^2$),   200g/cm$^2$
thickness of the material is needed to stop the electromagnetic cascade.
The difference in the absorption length is used to discriminate 
isolated photons from hadron, which would be useful to study 
$\tilde{\tau}\rightarrow \tau\gamma X$ decay.
 
In a previous paper~\cite{Hamaguchi:2004df}, we discussed the
possibility to re-use of the existing 1 kton detector, such as SOUDAN
II~\cite{Allison:1996wn} as the CNLSP stopper.  The SOUDAN II consists
of the layers of $\mathcal{O}$(m) long drift tubes and thin iron
plates. The physics goal of this detector is the search for proton decays. 
To be sensitive to
the low energies involved in these decay processes,  
the average density of SOUDAN II is  low, less than 2g/cm$^3$.  The size 
of the detector  is probably
enough to stop a certain amount of CNLSPs and measure the decay rate,
and therefore it may well be  appropriate for the a first stage of the 
CNLSP
study. On the other hand, it is certainly not enough  for a 
detailed energy 
measurement
of the $\stau$ decay product. Most of the hadronic decay
cascades are not fully contained for a  detector with the geometry allowed
by the space in the CMS cavern. 
 A high
density detector consisting of the layers of drift tubes/scintillators or
RPCs to measure the charged particles between  iron plates thicker than
SOUDAN II will be more optimized for the CNLSP study.

In this paper we assume  a conservative  energy resolution for  hadronic
 showers, which  is around $150\% /\sqrt{E/{\rm GeV}}$.
  The value is not 
unrealistic  for a simple massive and 
affordable detector, if a shower is sufficiently 
contained in the stopper.
An additional complexity may occur for
showers which develop parallel to the layers of the 
tracking devices.  If the particles pass  mostly through tracking devices,  
they feel a much lower average density, while 
the particles going   mostly through the iron plates do not give 
detectable 
signature efficiently.  We assume that the measured energy will be
corrected depending on the shower directions.  A detector uniform to
all directions would be better to measure the energy of decay products
of the stopped CNLSP.  The calorimetery technologies studied for the ILC 
 may satisfy 
such conditions; see \cite{Martyn:2006as} for the CNLSP study at ILC.

\section{Supersymmetric Models with charged next lightest SUSY particles, and 
expected number of stopped CNLSP}
\label{sec:modelpoints}

In this section, we briefly describe supersymmetric models with a
long-lived charged next lightest SUSY particle.  We also select some
model points, and estimate at each model point the number of CNLSP that
can be stopped at the stopper--detector.

In the minimal supergravity (mSUGRA) models, the scalar masses and
gaugino masses, trilinear couplings are universal at the GUT scale
$M_\mathrm{GUT}$ which are denoted by $m_0$, $M_{1/2}$, and $A_0$
respectively.  The resulting mass spectrum is obtained by solving
renormalization group equations (RGE's). When the RGE is integrated up
to $\mathcal{O}(m_Z)$, slepton soft masses and gaugino masses are
approximately given by the following convenient formulas
\begin{eqnarray}
m^2_{\tilde{q}_L}
&\sim&
m_0^2+6.3 M^2_{1/2},\cr
m^2_{\tilde{u}_R} \simeq m^2_{\tilde{d}_R}
&\sim&
m_0^2+5.4 M^2_{1/2},\cr
m^2_{\tilde{\ell}_L}
&\sim&
m_0^2+0.5 M^2_{1/2},\cr
m^2_{\tilde{e}_R}
&\sim&
m_0^2+0.15 M^2_{1/2},\cr
\frac{M_i}{g^2_i}&=&\frac{M_{1/2}}{g_X^2},
\label{sugraeq}
\end{eqnarray}
where $g_X$ is the gauge couplings at the unification scale.\footnote{
The pole masses of strongly interacting SUSY particles receive a large
corrections of $\mathcal{O}$(30\%) if the mass scale is
$\mathcal{O}$(1TeV).}  
In this model, the mass of gravitino
$\mgravitino$ is order of $m_0$ or $M_{1/2}$, because both of them are
proportional to the $F_0/\Mpl$ with $\mathcal{O}(1)$ coefficient,
where $F_0$ is the fundamental SUSY breaking scale. Depending on the
$\mathcal{O}(1)$ coefficients, the gravitino can be the lightest
superparticle in general.  Also in the gaugino mediation
models~\cite{gauginoMSB}, the gravitino can be the LSP in a large
domain of the parameter space~\cite{Buchmuller:2005rt}.  In this case
the scalar masses are very small at the boundary, i.e., $m_0\simeq
0$, and the stau naturally becomes the NLSP.

In the gauge mediation (GM) model~\cite{Giudice:1998bp}
the supersymmetry is broken at lower
energy scale, and the SUSY breaking in the hidden sector is mediated
to the MSSM sector by gauge interactions.  The simplest GM model is
described by 6 parameters: $\Lambda=F/M_{\rm mes}$, $M_{\rm mes}$,
$N_5$, $\tan\beta$, sgn$\mu =\pm 1$ and gravitino mass.  Here $N_5$ is
an effective number of vector-like heavy quarks and leptons in SU(5)
representations, $\Phi$ and $\bar{\Phi}$. ($N_5=1$ for $5+\bar{5}$
quarks and leptons, and $N_5=3$ for $10+\overline{10}$.)  A messenger
field $Y$ couples to the vector-like fields as $W=\lambda
Y\Phi\bar{\Phi}$ and develops a vacuum expectation value $\lambda
\langle Y \rangle= M_{\rm mes}+\theta\theta F$.  The gravitino mass is
given by $\mgravitino = F_0/(\sqrt{3}\Mpl)$ where $F_0 (\ge F)$ is the
total SUSY breaking scale of the theory. We take $\mgravitino$ as a
free parameter in this model.

The masses of MSSM particles at $M_{\rm mes}$ are obtained by relatively
simple formula.  Gaugino masses satisfy the GUT relation, and 
\begin{eqnarray}
m_{\tilde{g}}&=&\frac{\alpha_s}{4\pi} N_5 \Lambda.
\end{eqnarray}
Squark and slepton masses are given by
\begin{eqnarray}
m^2_{\tilde{q}_L}
&=&
\left[
{8\over 3}
\left({\alpha_s\over 4\pi}\right)^2
+
{3\over 2}
\left({\alpha_2\over 4\pi}\right)^2
+
{1\over 30}
\left({\alpha_1\over 4\pi}\right)^2
\right] N_5 \Lambda^2,
\cr
m^2_{\tilde{u}_R}
&=&
\left[
{8\over 3}
\left({\alpha_s\over 4\pi}\right)^2
+
{8\over 15}
\left({\alpha_1\over 4\pi}\right)^2
\right] N_5 \Lambda^2,
\cr
m^2_{\tilde{d}_R}
&=&
\left[
{8\over 3}
\left({\alpha_s\over 4\pi}\right)^2
+
{2\over 15}
\left({\alpha_1\over 4\pi}\right)^2
\right] N_5 \Lambda^2,
\cr
m^2_{\tilde{\ell}_L}
&=&
\left[
{3\over 2}
\left({\alpha_2\over 4\pi}\right)^2
+
{3\over 10}
\left({\alpha_1\over 4\pi}\right)^2
\right] N_5 \Lambda^2,
\cr
m^2_{\tilde{e}_R}
&=&
\left[
{6\over 5}
\left({\alpha_1\over 4\pi}\right)^2
\right] N_5 \Lambda^2.
\label{gmeq}
\end{eqnarray}
 
All of the above models predict a large mass hierarchy between strongly
interacting superpartners and weakly interacting superpartners. 
Heavy gluino and squarks are copiously produced at the LHC, and 
they decay into the light weakly interacting SUSY particles. 
The lightest SUSY particle in the MSSM sector is 
either the lighter stau $\tilde{\tau}_1$ or the lightest neutralino 
$\tilde{\chi}^0_1$. 

The mass of the $\tilde{\tau}_1$ is the smaller eigenvalue of the
mass matrix,
\begin{equation}
{\cal M}^2=\left(
\begin{array}{cc}
m^2_{\tilde{\ell}_{L_3}}+m^2_{\tau}-\frac{1}{2}(2m_W^2-m_Z^2)\cos 2\beta & 
-m_{\tau}(A_{\tau}+\mu\tan\beta) \cr
-m_{\tau}(A_{\tau}+\mu\tan\beta)& 
m^2_{\tilde{\tau}_R}+m^2_{\tau}+(m_W^2-m_Z^2)\cos 2\beta  
\end{array}\right).
\end{equation}
Because of the off-diagonal elements of $\stau$ mass 
matrix, $\tilde{\tau}_1$ could be significantly lighter than 
the other sleptons.  If the $\tilde{\tau}_1$ is the NLSP, 
the stopper--detector is useful to stop it and to study its decay.
We therefore consider the phenomenology when $m_{\tilde{g}},
m_{\tilde{q}}\gg m_{\tilde{\chi}^0_1}>m_{\tilde{\tau}_1}$ in this paper. 
In the following, we omit the subscript of $\tilde{\tau}_1$ and 
denote the NLSP stau as $\stau$.

In most part of this paper, we will discuss the CNLSP physics 
as model independent as possible. However, 
to give some numerical reference,  we choose a few model 
points.  For mSUGRA models, we take the points proposed in 
\cite{albert}.  The parameters are listed in Table.~\ref{masssugra}
and mass spectrum is given by the ISAJET ver. 7.69. In \cite{albert}, 
the gravitino mass is taken to be $\mgravitino=m_0$ in those model points.

\begin{table}[h]
\begin{tabular}{|c||ccc|}
\hline 
Point & $\epsilon$ & $\zeta$ & $\eta$ \cr
\hline
$M_{1/2}$ & 440 & 1000 & 1000 \cr
$m_0$ & 20 & 100 & 20 \cr
$\tan\beta$ & 15 & 21.5 & 23.7\cr
\hline
$\mgl$[GeV]           &1025 & 2191 & 2190 \cr
$\mchii$[GeV]        & 175 & 417& 416\cr
$m_{\tilde{\tau}_1}$[GeV]        & 154.2& 343.5 & 324.3 \cr
\hline
$\sigma$(SUSY)[pb]& $3.03$ & $2.27\times 10^{-2}$ & $2.34 \times 10^{-2}$ \cr
stopped in the stopper--detector per $10^5$ events &255& 250 & 254\cr 
stopped for 300 fb$^{-1}$ & 4636 & 34& 36 \cr
\hline 
\end{tabular}
\caption{Some model points in mSUGRA model from \cite{albert}. 
 The mass spectrum and production 
cross section relevant to our study are shown.  }
\label{masssugra}
\end{table}

\begin{table}[h]
\begin{tabular}{|c||ccccc|}
\hline 
$\Lambda$[TeV]& 40&  50& 60& 70& 80\cr 
$\mgl$[TeV]& 0.93 & 1.13 & 1.34 & 1.54 & 1.74 \cr 
$\mchii$[GeV]&  161.7 & 205.3 & 248.7 & 292.1 & 335.4\cr
$m_{\tilde{\tau}_1}$[GeV]& 120.5& 150.1 & 179.9 & 209.8& 239.9\cr
\hline
$\sigma$(SUSY)[pb]&  5.24& 1.68 & 0.64 & 0.28& 0.13\cr   
stopped in the stopper--detector per $10^5$ events & 282&274& 274& 294& 302\cr
stopped for 300fb$^{-1}$  & 8830 &2762& 1052 & 494 & 236 \cr 
\hline
\end{tabular}\caption{Some model points in gauge 
mediation model. The production cross section and mass spectrum relevant in
 this study are also shown.}
\label{mass}
\end{table}

For the gauge mediation model, we take the model points similar to
that for the study in~\cite{ATLASTDR}.  Namely, we fix $N_5=3$,
$\tan\beta=15$ and $\Lambda/M_{\rm mes}=0.5$, where $\Lambda=40,50,60,
70, 80$~TeV.  The mass spectrum and production cross sections are
summarized in Table. \ref{mass}.  The gravitino mass for these model
can be very small, whose minimum value is given by
$\mathcal{O}(\Lambda^2)/\Mpl$.  If the messenger sector couples a
fraction of total SUSY breaking, i.e. $F < F_0$, the gravitino mass,
which is proportional total SUSY breaking $F_0$, can be large.

The pattern of the mass spectra are similar for both mSUGRA 
and GM sample points. 
The gaugino masses obey the GUT 
relation, and $\stau$ mass is lighter than $\tilde{B}$  mass because 
of the CNLSP assumption.  For mSUGRA, this implies $m_0<M_{1/2}$, 
so that  $\tilde{q}$ mass is about as large as $\tilde{g}$, as it is so in
GM  models. The relation $m_{\tilde{q}}\sim m_{\tilde{g}} \gg \mstau$ is 
realized in our model points. 

To estimate the number of stopped CNLSP, 
the  production cross section of the SUSY particles 
and the velocity distribution of $\stau$
at LHC must be evaluated.  They depend on the $\tilde{g}$ and $\tilde{q}$ masses, and
the mass difference $m_{\tilde{g}(\tilde{q})}-\mstau$.   We 
estimate the production cross section and the decay distribution 
by using HERWIG\cite{HERWIG}, where the mass spectrum and branching ratios 
are interfaced from ISAJET\cite{ISAJET}  to HERWIG by using ISAWIG\cite{ISAWIG}. 
We generated $10^5$ SUSY events for  each model point. 

The flying range $R$ of the charged stable massive particle may be
calculated by integrating the energy loss equation of heavy ionizing
particle (Bethe-Bloch equation).  The result is a function of $\beta=
p/E$, with a linear dependence on the mass of the particle $M$. In
this paper, the stopping power of the stopper--detector is calculated
using the data in \cite{ICRU}.  See also the detailed discussion
in Ref.~\cite{Feng:2004yi}.

\begin{figure}[t]
\begin{center}
\includegraphics[width=12cm]{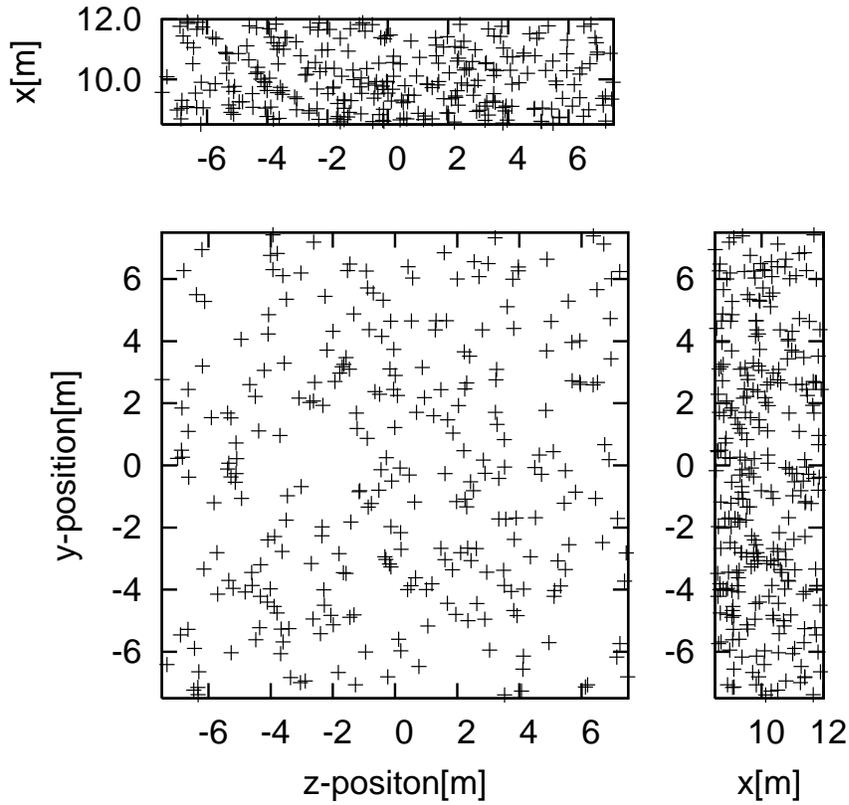}
\end{center}
\caption{The simulated positions of the CNLSPs in a stopper--detector
for the $\Lambda=40$~TeV GM point.  Here the $z$ axis
is the beam direction, and the $y$ axis is the vertical direction.  The
origin $(0,0,0)$  is the interaction point and
we assume the stopper--detector is at 8.5m$<x<$12m, $-7.5$m$<y<$7.5m, 
and $-7.5$m$<z<$7.5m (cf. Fig.~\ref{Fig:stopper}). 
The big square is the projection on the
$y$-$z$ plane, the top rectangle is the projection on the $x$-$z$ plane, 
and the right rectangle is the projection on the $x$-$y$ plane.  }
\label{Fig:position}
\end{figure}

The maximum length of a particle track through the stopper--detector
$l$(max), which depends on CNLSP direction, is calculated assuming
that the track of the CNLSP goes straight from the production
point. We regard the particle is stopped in the stopper--detector if
the flying range $R$ satisfies:
\begin{equation}
  \frac{2500~\mathrm{g/cm^2}}{\sin\theta}
  \;<\;
  R
  \;<\;
  \frac{2500~\mathrm{g/cm^2}}{\sin\theta}
  +
  l{\rm (max)}\times \rho_{\rm stop}\;,
\end{equation}
where $\rho_{\rm stop}=5$g$/$cm$^3$ is the density of the
stopper--detector. The number of stopped CNLSP in a stopper per $10^5$
events, and the number of stopped CNLSP for ${\cal L}= 300{\rm
fb}^{-1}$ in the two stoppers,  are listed in Table \ref{masssugra} and
\ref{mass}.  In Fig.~\ref{Fig:position} we also show the simulated
stopped positions in a stopper--detector.  The position distribution
is uniform in the detector.

\begin{figure}[t]
\begin{center}
\includegraphics[width=7cm, angle=90]{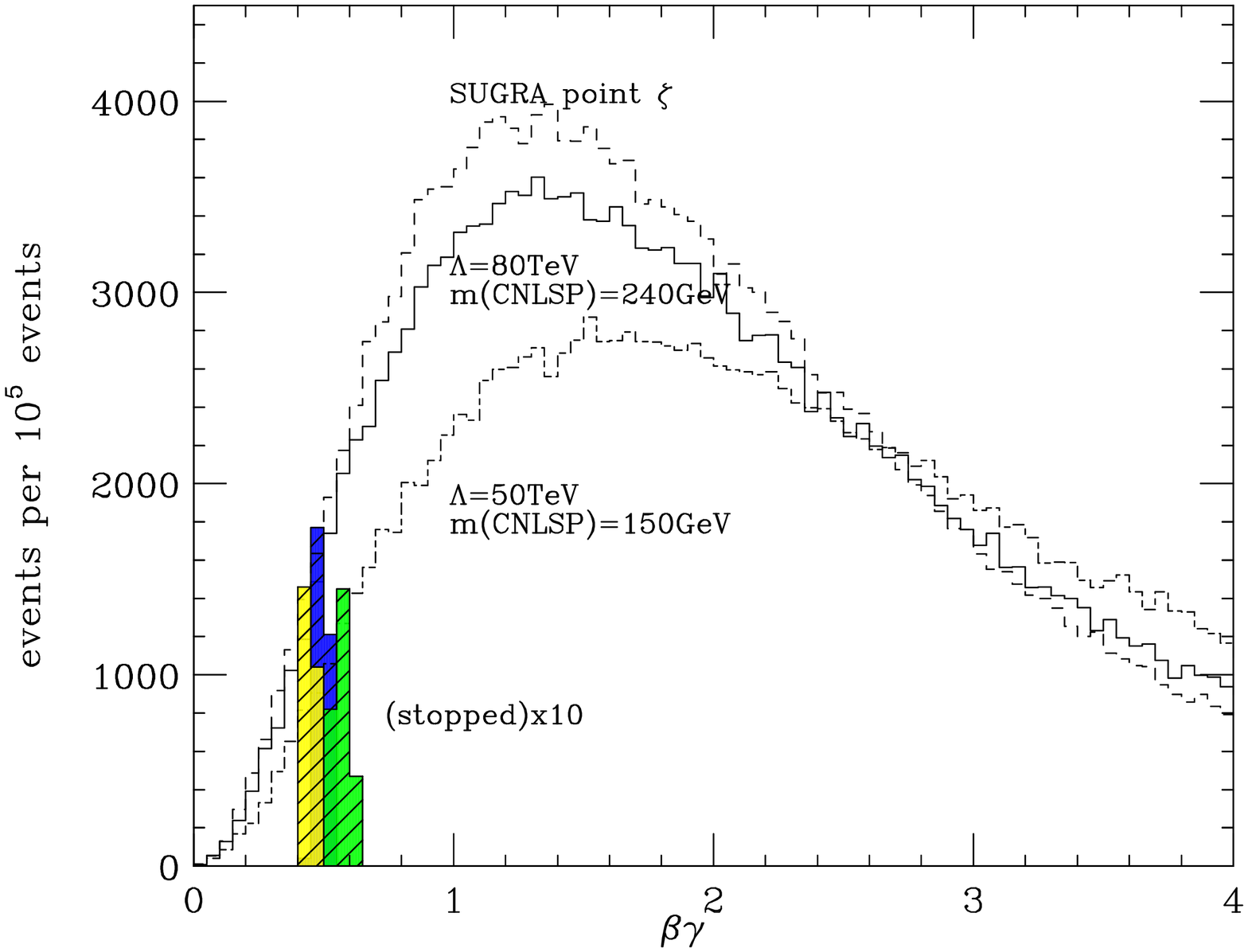}
\end{center}
\caption{The $\beta\gamma$ distribution of $\stau$ for $\Lambda=80$~
TeV and $\Lambda=50$~TeV in GM models and for point $\zeta$ in mSUGRA
models for $10^5$ SUSY events.  
Light (dark, gray) shaded histograms are the number of stopped events in the 
stopper scaled by factor of 10, for point $\zeta$ and for GM models with $\Lambda=80, 50$ TeV 
respectively. 
}
\label{Fig:betagamma}
\end{figure}

In Fig.~\ref{Fig:betagamma}, we show 
the $\beta\gamma$  distribution of the CNLSPs  for a few  model points. 
We find that CNLSP tends to be less relativistic when the SUSY 
scale is large, because momentum of parent squarks and gluino is reduced.  
In Fig.~\ref{Fig:betagamma}, the peak position 
of the $\beta\gamma$  is at $\sim 1.5(1.3)$ 
for $\Lambda=50(80)$~TeV.  
Because of that, the number of
CNLSP in the smaller  $\beta\gamma$ region 
is increased as the SUSY scale is increased. 
On the other hand, as $m_\mathrm{CNSLP}$ increases, 
CNLSPs with  smaller $\beta\gamma$ are stopped in 
the detector while
the number of events in the smaller $\beta\gamma$ bins are 
kinetically suppressed.  
For instance, for $\Lambda=50$ and 80~TeV,  CNLSPs in  
the bins between $0.5<\beta\gamma<0.6$ and 
$0.45<\beta\gamma<0.55$ are stopped in the stopper--detector, 
respectively.
Altogether, the number of 
stopped $\stau$ for $\Lambda=40$ to $80$~TeV 
for $10^5$ SUSY events is roughly constant as we can 
see in Table. \ref{mass}. 

The production cross section reduces when the gluino mass is increased
because the parton distribution functions of gluon and quarks are
quite small for $x\gg 0.1$.  For 300fb$^{-1}$, the number of events
stopped at the assumed two 4 kton stoppers ranges from 8000 events to
30 events in the table.  We will see in the next section that
accumulation of $\mathcal{O}(100)$ CNLSPs are enough to measure the
lifetime with $\mathcal{O}$(10)\% accuracy.  We will also estimate the
resolution of $\tau$ lepton energy $E_{\tau}$ arising from the decay
$\stau\to \tau X$ through the end point of the tau jet energy.
Statistically $\mathcal{O}(1000)$ stopped CNLSPs are enough to measure
the end point with a few GeV error.  In the upgrade of LHC (SLHC),
integrated luminosity of 3000fb$^{-1}$ is proposed, therefore the
number of stopped events ranges from $\mathcal{O}(300)$ to
$\mathcal{O}(80000)$ for the model points presented in
Table~\ref{mass}.

\section{Study of the $\stau$ two body decay in stopper--detector}
\label{sec:twobody}

\begin{figure}[t]
\begin{center}
\includegraphics{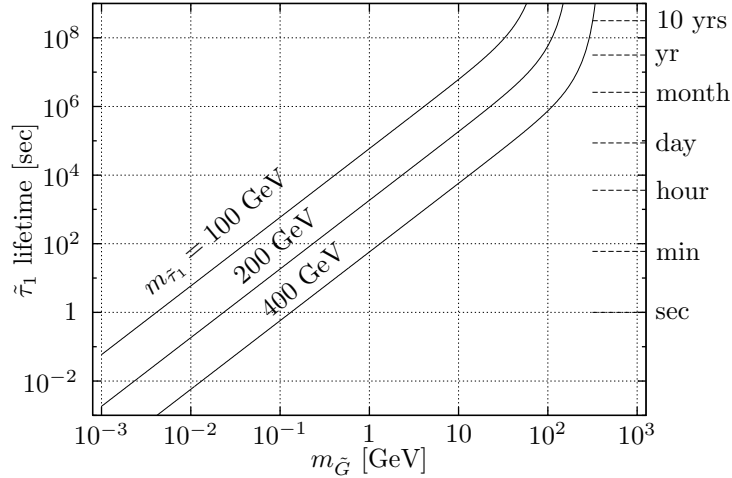}
\end{center}
\caption{The lifetime of the CNLSP $\stau$ in the case of gravitino
LSP, as a function of the gravitino mass, for the stau mass 100, 200,
and 400~GeV (from top to bottom).}
\label{Fig:lifetime}
\end{figure}

In this section we study the two body decay of the CNLSP in the
stopper--detector.  Both in mSUGRA and GM models, the stau can be the
CNLSP and decays into the gravitino $\gravitino$ and the $\tau$--lepton.
The CNLSP decay width into a gravitino and a lepton 
is given by~\cite{Buchmuller:2004rq}
\begin{eqnarray}
  \Gamma_{\stau}(\stau\to\gravitino\tau)
  &=&
  \frac{\mstau^5}
       {48\pi \mgravitino^2 \Mpl^2}
       \left(1-
       \frac{\mgravitino^2 + m_\tau^2}
	    {\mstau^2}
	    \right)^4
	    \left[
	    1-\frac{4 \mgravitino^2 m_\tau^2}
	    {(\mstau^2-\mgravitino^2-m_\tau^2)^2}
	    \right]^{3/2}\;.
	    \nonumber\\
	    &=&
	    (68~\mathrm{days})^{-1}
	    \left(\frac{\mstau}{100~\mathrm{GeV}}\right)^5
	    \left(\frac{10~\mathrm{GeV}}{\mgravitino}\right)^2
	    \times 
	    \nonumber\\
	    &&
	    \left(1-
	    \frac{\mgravitino^2 + m_\tau^2}
		 {\mstau^2}
		 \right)^4
		 \left[
		   1-\frac{4 \mgravitino^2 m_\tau^2}
		   {(\mstau^2-\mgravitino^2-m_\tau^2)^2}
		   \right]^{3/2}\;.
		 \label{eq:Gamma32}
\end{eqnarray}
We show the dependence of the stau lifetime on the gravitino mass in
Fig.~\ref{Fig:lifetime}.

In general, the two--body decay $\stau\rightarrow X \tau $ can be
triggered by a single tau jets initiating from the position where
$\stau$ is stopped.  The tau energy is monochromatic and expressed as
\begin{equation}
  E_{\tau}=\frac{\mstau^2 + m^2_{\tau} - m^2_X}{2 \mstau}\;.
\label{etau}
\end{equation}
Here $X$ is the invisible particle in the $\stau$ decay, in this case
$X=\gravitino$.  If one can measure both of the lifetime and the mass
of the stau, the gravitino mass can be determined {\it assuming} that
Eq.(\ref{eq:Gamma32}) is correct.  Then the total SUSY breaking scale
$F_0=\sqrt{3} \mgravitino \Mpl$ is also determined, which is very
important to understand the hidden sector physics.

The LHC main detectors can determine the mass of $\stau$ through the
stau velocity measurement $\beta_{\stau}$, e.g., in the muon system of the
CMS detector. However, the measurement of the lifetime may not be easy
at the main detectors if the lifetime is too much longer than the
detector size.  The cross section is typically $\mathcal{O}(1)$ pb or
less, therefore we have at most $\mathcal{O}(10^5)$ event at hand for
${\cal L}= 100{\rm fb}^{-1}$. Thus, only 10 events or less decay
inside the detector if $c\tau>100$ km. 
Some of the CNLSP are stopped in the main detector, but measuring the
decay precisely in the main detector would be challenging during the
beam time.  On the other hand a stopper--detector~\cite{Hamaguchi:2004df} 
can measure the
position and the time where a CNLSP is stopped, and its decays without
dead-time. The lifetime measurement will be discussed in
Sec.\ref{subsec:lifetime}. See \cite{albert} also on the idea to
measure the lifetime by triggering muons from the decays of the CNLSP
stopped in the surrounding rock.

\begin{figure}[t]
\begin{center}
\includegraphics{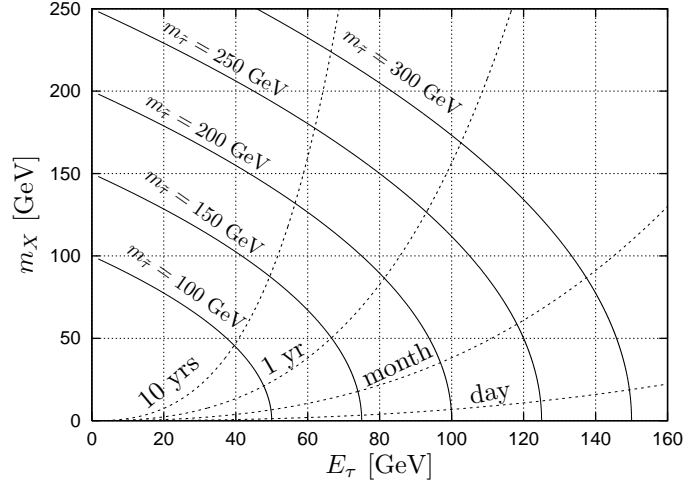}
\end{center}
\caption{Sensitivity of the reconstructed LSP mass $m_X$ [see
Eq.(\ref{eq:mX})] on the tau energy $E_\tau$, for $\mstau=100$, 150,
200, 250, and 300~GeV (solid lines). Dashed lines show contours of the
lifetime in the case of the gravitino LSP ($X = \gravitino$).  }
\label{Fig:Etau}
\end{figure}

To {\it predict} the CNLSP lifetime, one has to determine the
gravitino mass independently. This is possible through the extraction
of $E_{\tau}$ from the energy distribution of the tau jet from the
CNLSP decay, because $E_{\tau}$ is a function of $\mstau$ and $m_X$,
as can be seen in Eq. (\ref{etau}). $m_X$ is expressed as a function
of $E_{\tau}$ as follows;
\begin{eqnarray}
  \label{eq:mX}
  m_X
  &=& \sqrt{\mstau^2 + m_\tau^2 - 2 \mstau E_\tau}\;.
\end{eqnarray}
Fig.~\ref{Fig:Etau} shows the dependence of the reconstructed LSP mass
Eq.(\ref{eq:mX}) on the tau energy $E_\tau$ for several values of
$\mstau$.  As can be seen from the figure, it is crucial to measure
the $E_\tau$ as precise as possible, especially for small $m_X$, in
order to determine the mass $m_X$.
Hence, the stopper--detector should offer a reliable measurement of
$E_{\tau}$.  We will discuss the $E_{\tau} $ measurement and the $m_X$
reconstruction in Sec.~\ref{subsec:Etau} and Sec.~\ref{subsec:mX},
respectively.

One can study the supersymmetric version of gravity interaction 
by studying the consistency between the observation and prediction of the decay rate. 
If the gravitino is the LSP, the decay width is given by
Eq.(\ref{eq:Gamma32}). Now if one can independently determine the
gravitino mass by means of kinematics, as described above,
Eq.(\ref{eq:Gamma32}) can be used in the other way round, which leads
to the measurement of the 'supergravity Planck
scale'~\cite{Buchmuller:2004rq}
\begin{eqnarray}
  \Mpl^2({\rm supergravity}) &=&
  \frac{\mstau^5}
       {48\pi \mgravitino^2 \Gamma_{\stau}}
       \left(1-
       \frac{\mgravitino^2+m_\tau^2}
	    {\mstau^2}
	    \right)^4
	    \left[
	    1-\frac{4 \mgravitino^2 m_\tau^2}
	    {(\mstau^2-\mgravitino^2-m_\tau^2)^2}
	    \right]^{3/2}\;.
       \label{eq:Mp}
\end{eqnarray}
Comparison of the obtained $\Mpl({\rm supergravity})$ with the Planck
scale of the Einstein gravity $\Mpl({\rm gravity})=(8\pi
G)^{-1/2}=2.43534(18)\times
10^{18}~\mathrm{GeV}$~\cite{Eidelman:2004wy} would be a crucial test
of the supergravity. Prospects of the ``Planck scale'' measurement will be
discussed in Sec.\ref{subsec:Mpl}.

It should be noted that the undetectable particle $X$ may not be the
gravitino $\gravitino$.  Any particle which couples weakly to $\stau$
can be particle $X$.  If the decay width into $X\tau$ is larger than
the decay width into $\gravitino\tau$, i.e.  $ \Gamma(\stau
\rightarrow \tau X)> \Gamma(\stau \rightarrow \tau \gravitino)
$,\footnote{This includes the trivial case 
where the decay into the gravitino
is kinetically forbidden, $\Gamma(\stau\to\tau\gravitino)=0$, i.e.,
$\mgravitino > \mstau >m_X$. } the CNLSP lifetime may be different
from the supergravity prediction obtained from measured $\mstau$ and tau jet
energy distribution.  Inconsistency between the measured and predicted
lifetime immediately means a discovery of a new sector that may not
be accessible otherwise.

One of well motivated examples of such a non-SUGRA decay is $\stau
\rightarrow \axino \tau$ where $\axino$ is the axino, superpartner of
the axion.  The CNLSP $\stau$ decay into axino is studied in
Ref.~\cite{Brandenburg:2005he} for hadronic, or KSVZ axion
models~\cite{Kim:1979if+X}. In this paper we adopt the set--up in
Ref.~\cite{Brandenburg:2005he} for the axino interaction, which we
briefly describe here.

In KSVZ axion model, the Peccei-Quinn (PQ)
mechanism~\cite{Peccei:1977hh+X} is realized in an action with new
heavy quarks. When these heavy quarks are integrated out, anomalous
terms involving the axion and gauge bosons are generated at low energy
effective action.
When the axion interaction is supersymmetrized, its fermionic
superpartner, the axino $\axino$, must be introduced. The axino mass
$\maxino$ can range between the eV and GeV scale depending on the
model and SUSY breaking scheme~\cite{axinomass}, and we regard it as a
free parameter in this paper. The coupling of the axino to the bino
and the photon/$Z$-boson at the scale below the Peccei-Quinn scale
$f_a$ is given by the Lagrangian
\begin{equation}
\label{eq:AxinoBino}
L_{\axino}=i\frac{\alpha C_{aYY}}{16\pi \cos\theta_W^2 f_a}\bar{\axino} 
\gamma_5[\gamma_{\mu},\gamma_{\nu}]\tilde{B}(\cos\theta_WF_{\mu\nu}
-\sin\theta_WZ_{\mu\nu}). 
\end{equation}

\begin{figure}[t]
\begin{center}
\includegraphics[width=9cm]{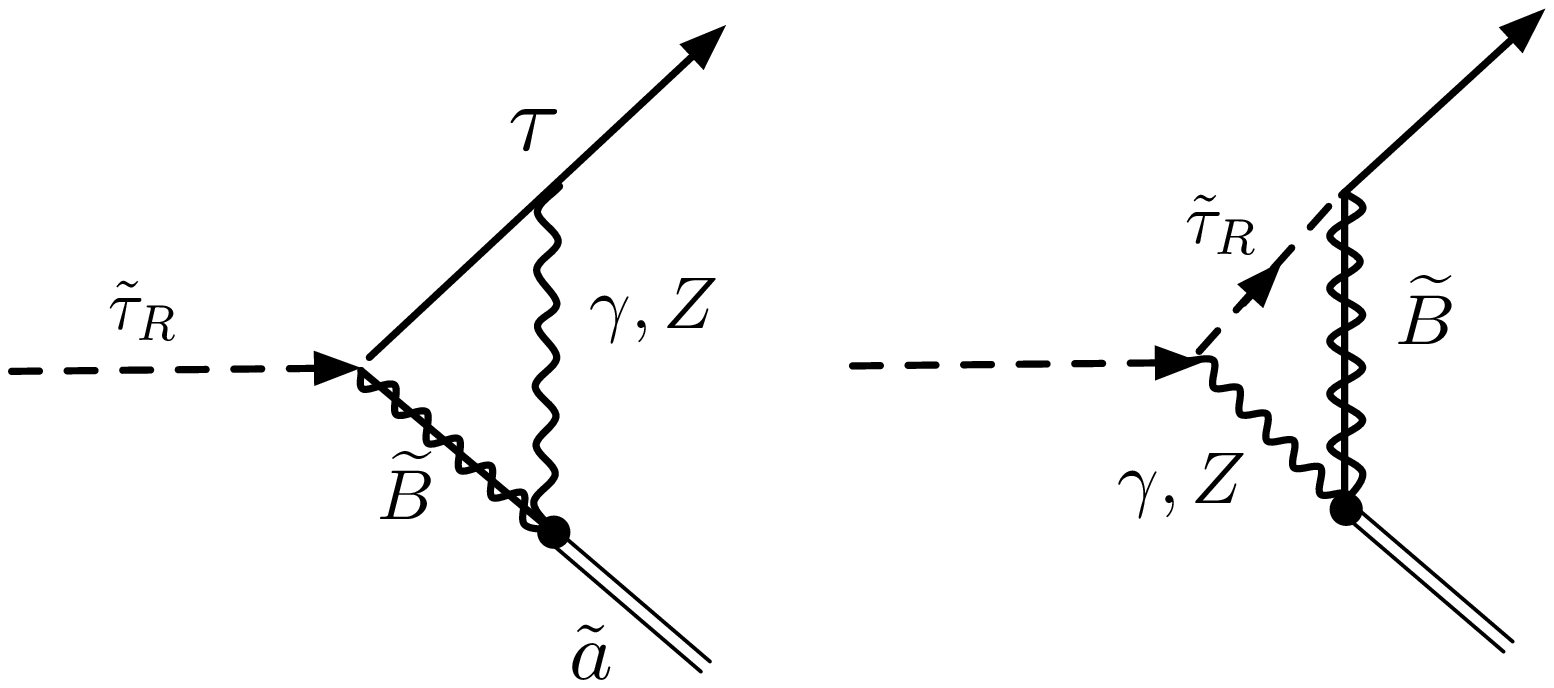}
\end{center}
\caption{
Feynman diagrams for $\stau\rightarrow\axino\tau$.
}
\label{Fig:twobody}
\end{figure}
The action does not contain direct $\stau\tau\axino$ coupling and also
strongly suppressed by the PQ scale $10^{9}{\rm GeV }\lsim f_a\lsim
10^{11}{\rm GeV}$.  The two body decay $\stau\rightarrow \axino\tau$
is induced by the one loop diagram shown in Fig.\ref{Fig:twobody}.
The loop integral has a logarithmic divergence. This is because the
effective vertex (\ref{eq:AxinoBino}) is applicable only if the
momentum is smaller than the heavy (s)quark masses, whereas the loop
momentum exceeds that scale.  In the full theory, one calculates
two--loop diagrams with the heavy (s)quarks, which leads to a finite
result~\cite{axino2loop}. Here we regulate the logarithmic divergence
with the cut--off $\sim f_a$~\cite{Brandenburg:2005he,Covi:2002vw}, 
and the effective 
$\axino\tau\tilde{\tau}_R$ coupling is parameterized as
\begin{equation}
L\simeq -\xi C_{aYY} \frac{3\sqrt{2}\alpha^2}{8\pi^2 \cos^4\theta^4_W}
\frac{m_{\tilde{B}}}{f_a}
\log\frac{f_a}{m} \tilde{\tau}_R \bar{\tau}P_L\axino
+hc, 
\label{eq:AxinoStau}
\end{equation}
Here, $m\simeq m_{\tilde{\tau},\tilde{B}}\simeq
\mathcal{O}(100~\mathrm{GeV})$ and we take $\log(f_a/m)=20.7$ 
hereafter. The parameter $\xi$ is an order one parameter to represent
the uncertainty coming from the cut--off procedure mentioned above. In
this paper, we regard this as a free parameter.  The two body decay
width is given as~\cite{Brandenburg:2005he}
\begin{eqnarray}
\Gamma( \stau\rightarrow \axino\tau )
&=&\frac{9\alpha^4C^2_{aYY}}
   {512\pi^5\cos^8\theta_W}
\frac{m^2_{\tilde{B}}}{f^2_a}
\frac{(\mstau^2 -\maxino^2)^2}{\mstau^3}
\xi^2\log^2 \left(\frac{f_a}{m}\right)
\cr
&=&\xi^2 (25~\mathrm{sec})^{-1} 
C^2_{aYY}
\left(   1-\frac{\maxino^2}{\mstau^2} \right)^2
\left(\frac{\mstau}{\rm 100GeV}\right)
\left(\frac{10^{11}{\rm GeV} }{f_a}\right)^2
\left(\frac{m_{\tilde{B}}}{\rm 100GeV}
\right)^2. 
\label{eq:axinoGamma}
\end{eqnarray}

\subsection{Lifetime measurement}
\label{subsec:lifetime}
In this subsection, we estimate the statistical error in the CNLSP
lifetime measured by the stopper--detector.  The analysis is
model--independent and applicable to any long-lived charged particle
stopped in the stopper.

For each CNLSP stopped in the stopper--detector, the stopping time
$t_\mathrm{stop}$ and the decaying time $t_\mathrm{decay}$ will be
recorded. The lifetime is then measured by fitting the temporal
distribution of the decaying events $N(t)$, where $t =
t_\mathrm{decay} - t_\mathrm{stop}$. Here, we use a
maximum--likelihood fitting and adopt the following procedure: First
we define a time $t_e$ such that $N(t<t_e) =
(1-e^{-1})N_\mathrm{total}\simeq 0.632 N_\mathrm{total}$
and $N(t>t_e) =
e^{-1}N_\mathrm{total}\simeq 0.368 N_\mathrm{total}$, 
where
$N_\mathrm{total}$ is total number of stopped event.\footnote{More
precisely, $t_e$ is defined by $t_e=(t_j+t_{j+1})/2$, where
$j<(1-e^{-1})N_\mathrm{total}<j+1$ and $t_j$ is the decay time of the
$j$-th event: $t_1<t_2<\cdots <t_j<\cdots <t_{N_\mathrm{total}}$.}
(For large $N_\mathrm{total}$, this $t_e$ is already a good estimator
of the lifetime.)  We then calculate the $\ln L$ distribution as a
function of lifetime $\tau$:
  \begin{eqnarray}
    \ln L (\tau) &=& \sum_{i = \mathrm{bins}}
      f_\mathrm{P}\left(n_i\;; \nu_i(\tau)\right)
  \end{eqnarray}
  where $f_\mathrm{P}(n;\nu)=\nu^n e^{-\nu}/n!$ is the Poisson
  distribution, $n_i$ is the number of events in the $i$-th bin, and
  $\nu_i(\tau)$ is the predicted average number of events in that bin:
  \begin{eqnarray}
    \nu_i(\tau) &=& N_\mathrm{total}\left(
    e^{-(i-1)\Delta t/\tau}
    -e^{-i\Delta t/\tau}
      \right).
  \end{eqnarray}
Here, we take 1 bin $=\Delta t = t_e/5$. 

\begin{figure}[t]
\begin{center}
\includegraphics{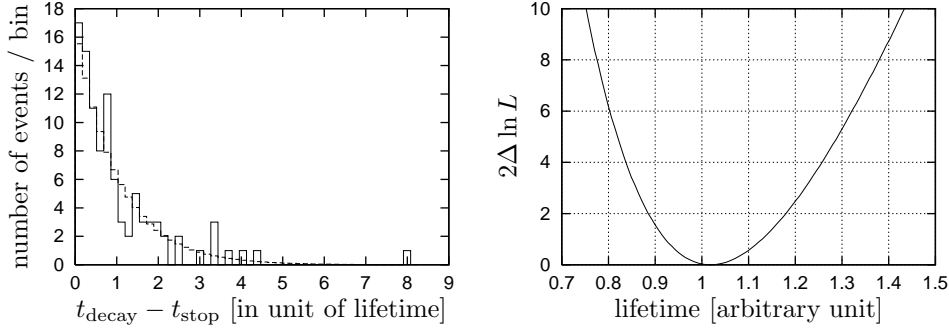}
\end{center}
\caption{An example of maximum--likelihood fit of the temporal
distribution $N(t)$ for $N_\mathrm{total}=100$. In the left plot, the
solid line is the generated events, and dashed line is the best fit.}
\label{Fig:lifetime_measure100}
\end{figure}

An example of the time distribution and the best fit curve for MC distribution 
are shown in
Fig.~\ref{Fig:lifetime_measure100} 
for $N_\mathrm{total}=100$.
The $n\sigma$ confidence interval can be estimated by the range
$2\Delta \ln L = 2(\ln L_\mathrm{max}-\ln L)\le n^2$. 
Fig.~\ref{Fig:lifetime_measure100} shows that
the error of the lifetime is $\Delta\tau/\tau = (10-15)\%$ for
$N_\mathrm{total}=100$.  We have done the same analysis 
for $N_\mathrm{total}=1000$ and found  $\Delta\tau/\tau = (3-4)\%$.

So far, we assumed that most of the stopped CNLSPs decay within the
experimental time scale.  We now consider the case where the lifetime
is longer. Suppose that one observes $N_\mathrm{total}=1000$ stopped
CNLSPs and only 10 events decaying within 1 year.  In such a case, the
lifetime is estimated from the number of the decaying events.  For
instance, 95\% interval of the mean $\nu$ of Poisson variable for
$n=10$ is $\nu = [5.4, 17.0]$~\cite{Eidelman:2004wy}. Using $\nu =
N_\mathrm{total}(1-e^{-1~\mathrm{year}/\tau})$, a 95\% interval $58 <
\tau < 184$ years is obtained. For much longer lifetime or smaller
$N_\mathrm{total}$, only a lower bound of the lifetime is obtained.

We have assumed that the background is negligible. The background from
cosmic neutrino and hard neutrino produced from the main detector
interaction point is small (cf.~\cite{Hamaguchi:2004df}), however
careful study on the accidental background is necessary when
statistics is low.  This is beyond of the scope of this paper.

\subsection{Measurement of the $\tau$ energy
from distribution of the $\tau$ jet energies}
\label{subsec:Etau}

In this subsection,  we estimate
the uncertainty of the tau energy determination. Then, in
Sec.\ref{subsec:mX} we discuss the kinematical reconstruction of the
LSP mass. Schematically, the procedure is as follows:
\begin{eqnarray}
  E_\mathrm{jet}\;\;\mathrm{distribution}
  \;\;\to\;\;
  E_\tau
  \;\;\to\;\;
  m_X.
\end{eqnarray}

When $\stau$ decays into $\tau$ and invisible particle $X$, the 
tau energy $E_{\tau}$ is monochromatic (see Eq.(\ref{etau})).
$E_\tau$ can be obtained by fitting  the $\tau$ jet energy
distribution :
\begin{eqnarray}
  \frac{dN}{dE_\mathrm{jet}}(E_\mathrm{jet};\;E_\tau)
  &\qquad\mathrm{where}&
  E_\mathrm{jet}
  \;=\;
  \sum_{i\;\ne\;\nu,\;\mu}^{\mathrm{decay\;products}}E_i\;.
\end{eqnarray}
Among the decay products of the $\tau$ lepton we omit the neutrinos
and muons.  In order to see the prospects of $E_\tau$ measurement with
a finite number of events, we generate the events from $\tau$ decay by
using the TAUOLA~\cite{TAUOLA}, and we perform a maximum--likelihood
fitting of low statistics (``experimental'') event sets by high statistics
(``theoretical'') distributions.

In practice, the observed jet energy distribution depends not only on
the $E_\tau$, but also on the detector resolution $\Delta
E_\mathrm{jet}$ and the tau polarization $P_\tau$:
\begin{eqnarray}
  \frac{dN}{dE_\mathrm{jet}}
  (E_\mathrm{jet};\;E_\tau,P_\tau,\Delta E_\mathrm{jet}).
\end{eqnarray}
To obtain theoretical predictions for
$dN/dE_\mathrm{jet}(E_\mathrm{jet};\,E_\tau,P_\tau,\Delta E_\mathrm{jet})$, 
we have generated tau decay events with high statistics run  of
TAUOLA for the parameter space $E_\tau =$ 30 --- 125~GeV
and $-1\le P_\tau\le 1$, while fixing 
$\Delta E_\mathrm{jet}/E_\mathrm{jet}=$ $150\%$ $/\sqrt{E_\mathrm{jet}/{\rm GeV}}$.\footnote{We
have generated $10^6$ events for each of
the parameter sets $E_\tau=$ 30, 31, 32,... 109, 125~GeV
and
 $P_\tau =\pm1$ (i.e., $96\times 2$ parameter sets $\times 10^6$ events),
and interpolated the distribution between those parameter points.}
For each single $\tau$ decay generated by TAUOLA, the
jet energy $E_\mathrm{jet}$ is calculated, and then smeared by a
Gaussian fluctuation with a variance $\sigma^2 = (\Delta
E_\mathrm{jet})^2$.
\begin{figure}[t]
\begin{center}
\includegraphics{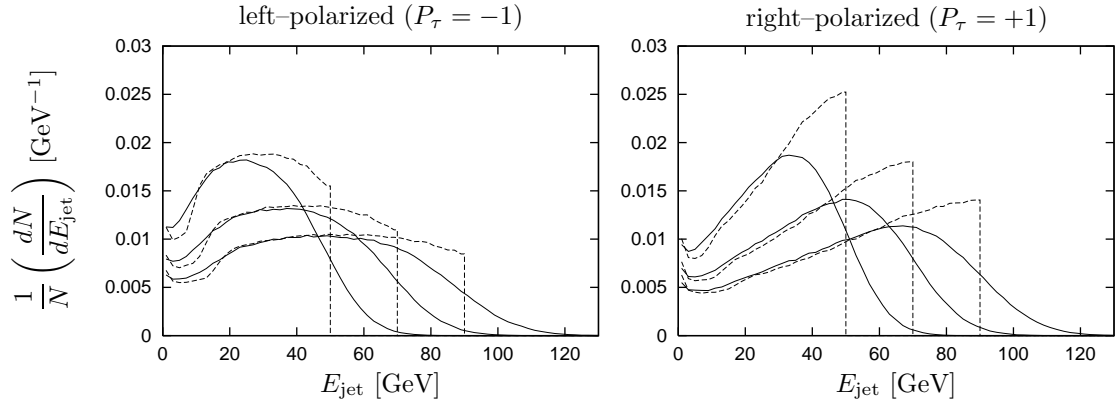}
\end{center}
\caption{Tau jet energy distributions for left--polarized (left) and
right--polarized (right) tau generated by TAUOLA. 
The primary tau energy is $E_\tau=50$,
70, 90~GeV from left to right.  The dashed lines show the spectrum
without energy resolution effect, and the solid lines show the
spectrum with detector energy resolution $\Delta
E_\mathrm{jet}/E_\mathrm{jet}$ $=150\%/\sqrt{E_\mathrm{jet}/{\rm GeV}}$.}
\label{Fig:taujet}
\end{figure}
Fig~\ref{Fig:taujet} shows examples of the jet energy distribution.

\begin{figure}[t!]
\begin{center}
\includegraphics{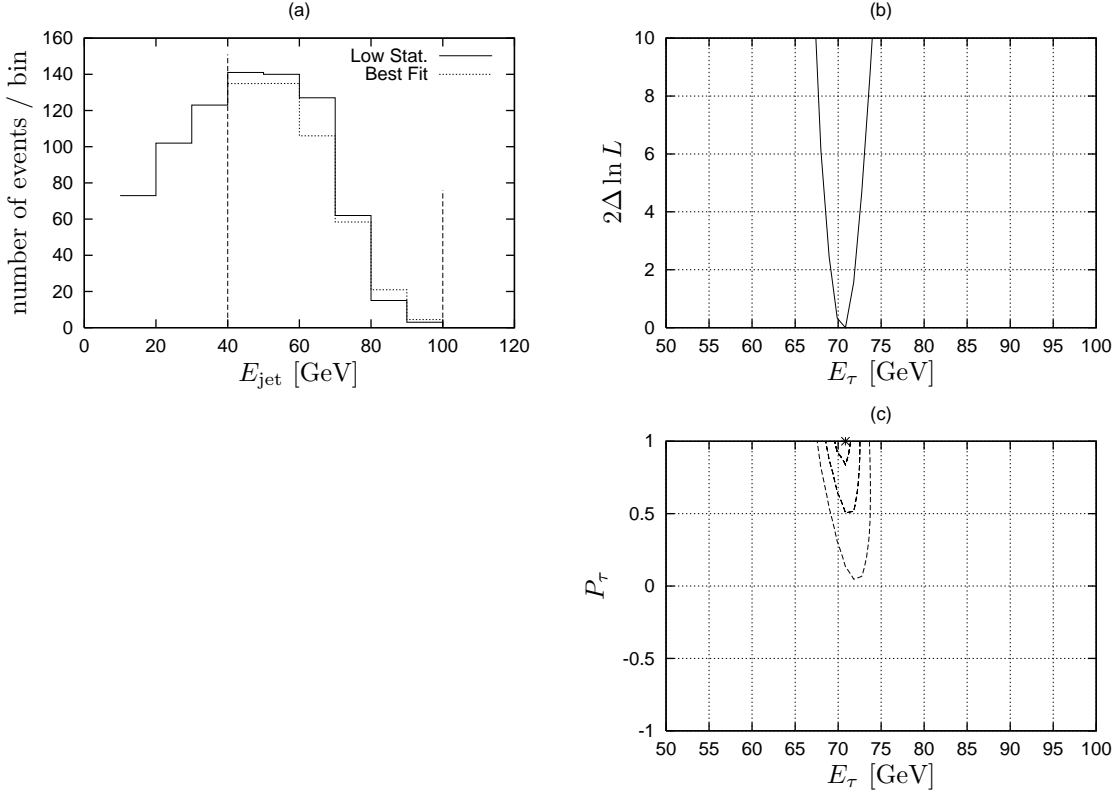}
\end{center}
\caption{An example of maximum--likelihood fit of low statistics
events from tau decays. (a) Energy distribution of tau jet events
generated from 1000 decaying tau leptons for $E_\tau=70$~GeV, $\Delta
E_\mathrm{jet}/E_\mathrm{jet} = 150\%/\sqrt{E_{\rm jet}/{\rm GeV}}$
and $P_\tau=+0.8$ (solid histogram), and the best fit distribution
(dotted histogram). Only the bins between the vertical lines are used
for the fit (see text). (b) $2\Delta \ln L =
2(\ln L_\mathrm{max} - \ln L)$ projected onto the $E_\tau$
axis. (c) Contour plots of $2\Delta \ln L = 1$, 4, 9 projected
onto the 
 ($P_\tau$, $E_\tau$)   
plane.}
\vspace{1em}
\label{Fig:lnLfit:1000}
\end{figure}

\begin{figure}[t!]
\begin{center}
\includegraphics{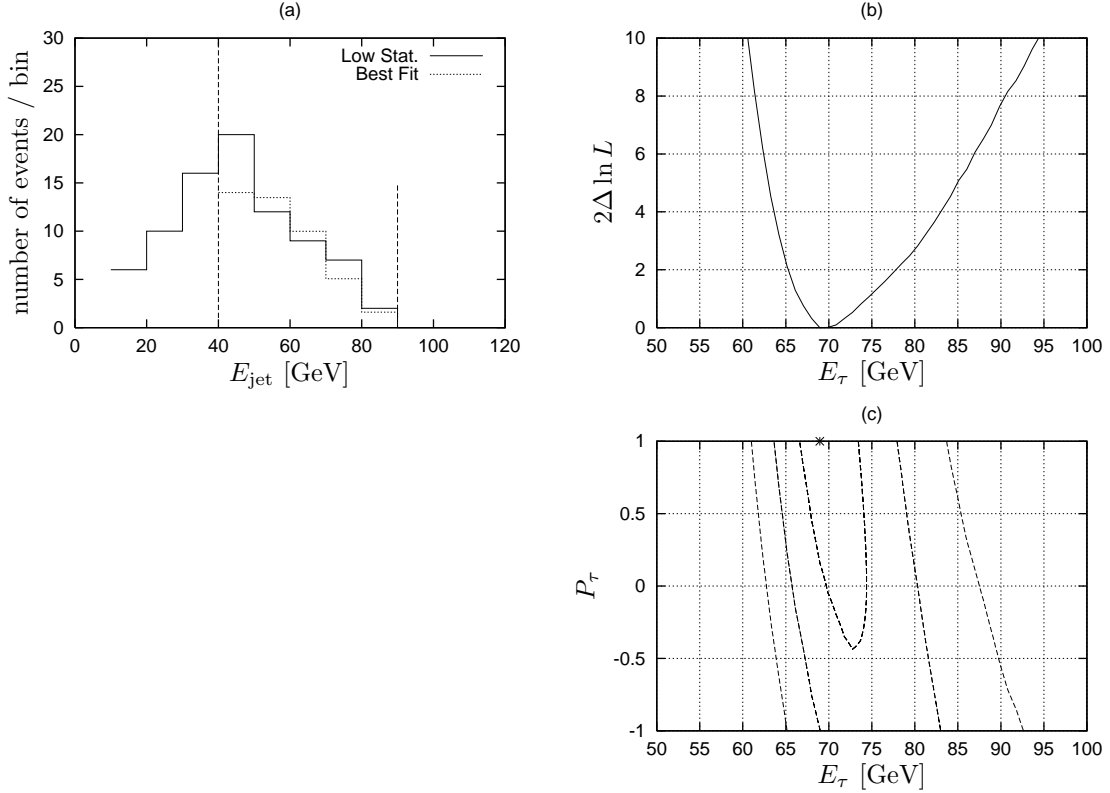}
\end{center}
\caption{An example of maximum--likelihood fit of low statistics
events from tau decays. The same as Fig.~\ref{Fig:lnLfit:1000} but
with 100 decaying tau leptons.}
\vspace{1em}
\label{Fig:lnLfit:100}
\end{figure}

In order to see a realistic situation, we generate a small number of
tau lepton decays for a fixed parameter set of ($E_\tau$, $P_\tau$, $\Delta E_\mathrm{jet}$), 
and then fit the result by the ``theoretical'' distribution obtained above. 
Here and hereafter, we assume that the energy resolution will be known in advance.
Examples are shown in
Fig.~\ref{Fig:lnLfit:1000} and Fig.~\ref{Fig:lnLfit:100} for total
number of decaying tau leptons $N_\tau=1000$ and $N_\tau=100$,
respectively. The energy of $\tau$ jets which are not contained in the detector 
may not be measured precisely. The $N_{\tau}$ corresponds to the 
number of well contained events. 
Some comments are in order here.  (i) We take 1 bin = 10~GeV. (ii) For
the maximum--likelihood fitting we take only the bins with number of
events $\ge 1$, and the bins above the peak energy, which in the
examples of Fig.~\ref{Fig:lnLfit:1000}(a) and \ref{Fig:lnLfit:100}(a)
correspond to the bins between the two vertical lines.  (iii) We then
calculate the $\ln L$ distribution in the parameter space of
($E_\tau$, $P_\tau$) as follows
\begin{eqnarray}
  \ln L (E_\tau, P_\tau) =
  \sum_{i=\mathrm{bins}} 
  \ln f_\mathrm{P}\left(N^\mathrm{low}_i\;;
  N^\mathrm{high}_i (E_\tau,P_\tau)
  \right)
\end{eqnarray}
where $f_\mathrm{P}(n;\nu)=\nu^n e^{-\nu}/n!$ is the Poisson
distribution, $N^\mathrm{low}_i$ is the number of events in the
$i$--th bin for the low statistics run, and
$N^\mathrm{high}_i(E_\tau,P_\tau)$ is the
predicted number of events in the $i$--th bin as a function of the
parameter set $(E_\tau,P_\tau)$, normalized by
the total number of decaying tau leptons.  Here, we assume that the
total number of stopped $\stau$ is known, i.e., the total number of
events is {\it not} taken as a free parameter for the fit.

The $n\sigma$ confidence interval can be estimated by the range
$2\Delta \ln L = 2(\ln L_\mathrm{max}-\ln L)\le n^2$ projected onto
the $E_\tau$ axis [see Figs.~\ref{Fig:lnLfit:1000}(b) and
\ref{Fig:lnLfit:100}(b)].  One can see that the primary tau energy can
be determined within an error of a few~GeV. From
Fig.~\ref{Fig:lnLfit:1000}(c) one can also see that the polarization
is hardly determined by the energy distribution analysis even with
$N_\tau = 1000$. This is because the sensitivity to the polarization
becomes very weak once the finite detector resolution is taken into
account (cf.  Fig.~\ref{Fig:taujet}).

\begin{figure}[t]
\begin{center}
\includegraphics{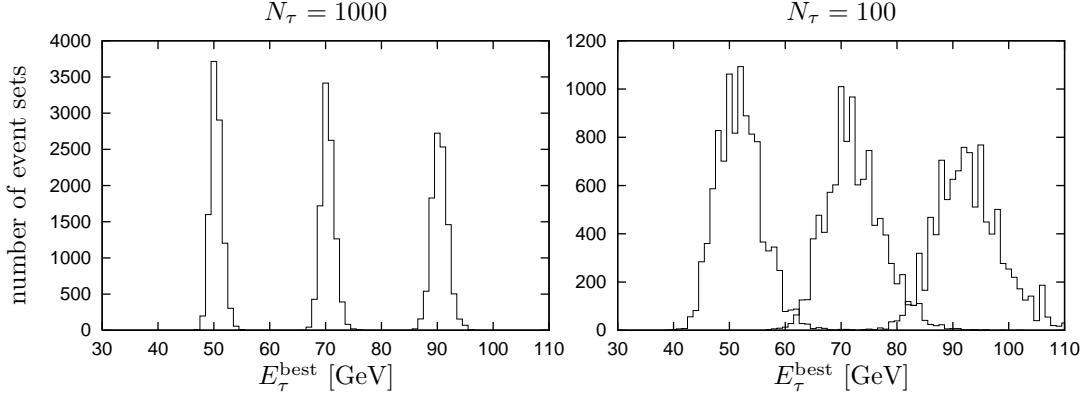}
\end{center}
\caption{Distribution of $E^{\rm best}_{\tau}$ for $E_\tau = 50$, 70 and 90 GeV
(from left to the right), for $N_\tau = 1000$ (left) and $N_\tau = 100$ (right).}
\label{Fig:delEtau}
\end{figure}

To estimate the statistical error in $E_{\tau}$ measurement 
we have generated the event sets with same statistics
$N_{\tau}=100$ $(1000)$  
and repeat the fit on $E_{\tau}$ and $P_{\tau}$ 
 to obtain the best fit value of  the  $E_{\tau}$, $E^{\rm best}_{\tau}$. 
In figure \ref{Fig:delEtau},   we show the 
distribution of $E^{\rm best}_{\tau}$ for  $E_{\tau}=50, 70, 90$~GeV,
$P_{\tau}=1$,  and  $\Delta E_\mathrm{jet}/E_\mathrm{jet}= 150\%/\sqrt{E/{\rm GeV}}$. 
{}From the variances of these distributions, we estimate the $1\sigma$ uncertainty of $E_\tau$ as
\begin{eqnarray}
\delta E_\tau/E_\tau 
&=&
61\% / \sqrt{E_\tau/\mathrm{GeV}}
\qquad {\rm for}\quad N = 100\;,
\\
\delta E_\tau/E_\tau 
&=&
15\% / \sqrt{E_\tau/\mathrm{GeV}}
\qquad {\rm for}\quad N = 1000\;.
\end{eqnarray}

\subsection{Determination of the LSP mass}
\label{subsec:mX}

\begin{figure}[t!]
\begin{center}
\includegraphics{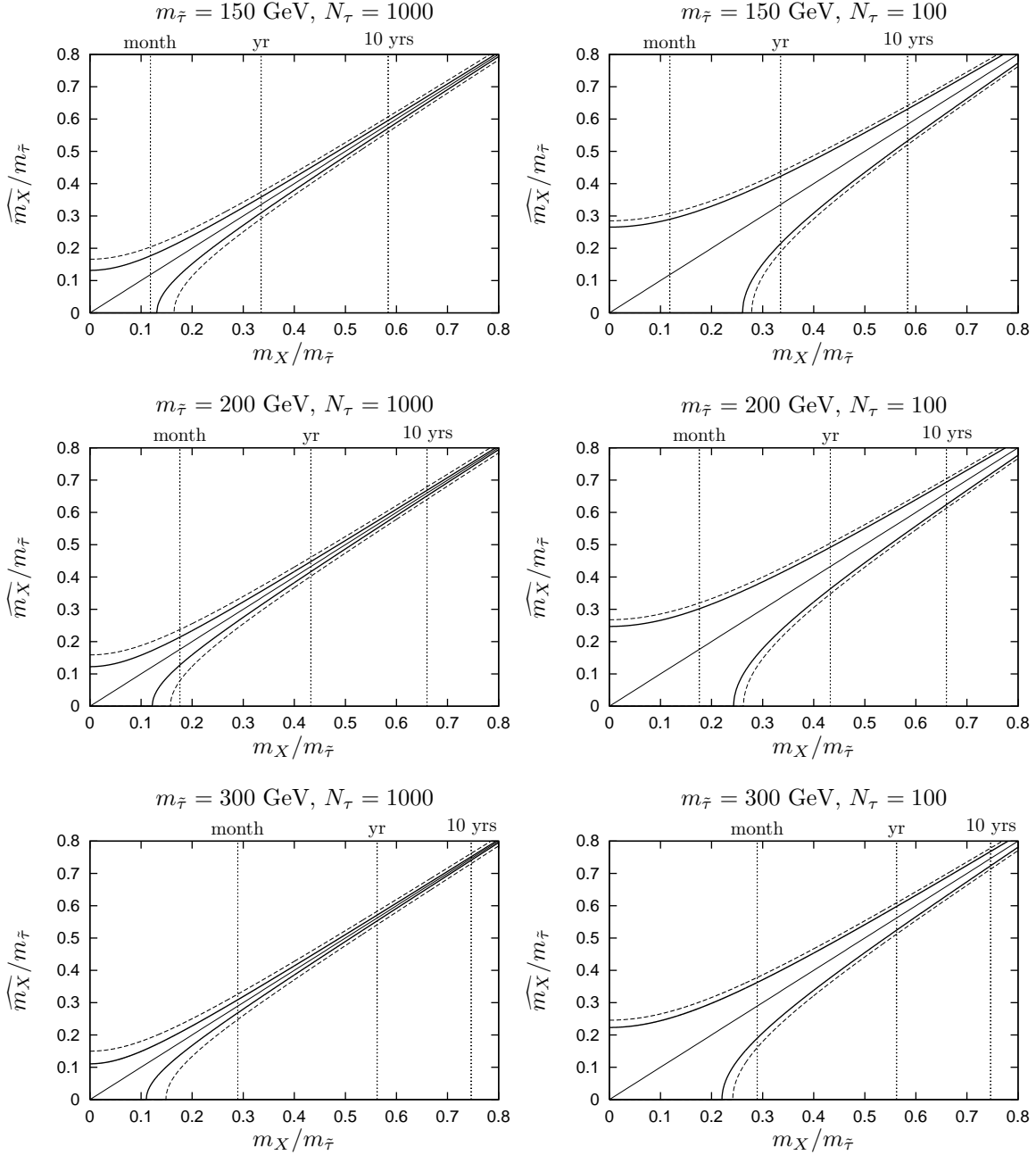}
\end{center}
\caption{Uncertainty of the reconstructed LSP mass, corresponding to
the estimated $1\sigma$ uncertainty of $E_\tau$, for $N_\tau = 1000$
and 100, and $\mstau = 150, 200$ and 300~GeV. Solid lines represent
the case without an error of $\mstau$, and dashed lines include 1\%
uncertainty of $\mstau$. Vertical dotted lines represent the stau
lifetime of 1 month, 1 year, and 10 years in the case of gravitino LSP
($X=\gravitino$).}
\label{Fig:mGmG}
\end{figure}

Now we can estimate the uncertainty of the LSP mass. In
Fig.~\ref{Fig:mGmG} we plot the range of reconstructed LSP mass
\begin{eqnarray}
  \widehat{m_X} &=& 
  \sqrt{\widehat{\mstau}^2 + m_\tau^2 - 2 \widehat{\mstau}
    \widehat{E_\tau}}\;,
\end{eqnarray}
where
\begin{eqnarray}
  E_\tau -  \delta E_\tau (E_\tau)
  \;\;\le\;\;
  \widehat{E_\tau}
  \;\;\le\;\;
  E_\tau + \delta E_\tau (E_\tau)
\end{eqnarray}
and $E_\tau = (\mstau^2 + m_\tau^2 - m_X^2)/ 2
\mstau$.
Note that the reconstructed LSP mass depends on not only $E_\tau$ but
also the measured CNLSP mass $\widehat{\mstau}$. In
Fig.~\ref{Fig:mGmG} we show the range of LSP mass for
$\widehat{\mstau}=\mstau$ (solid lines) and $0.99 \mstau\le
\widehat{\mstau}\le 1.01\mstau$ (dashed lines), the latter
corresponding to 1\% uncertainty of the CNLSP mass.
The stau mass determination from time of flight was discussed in 
\cite{ATLASTDR, Hinchliffe:1998ys, albert}.  
For CMS detector the mass resolution is estimated as 10--20\% in each event 
and less than 1\% for $\sim 1000$ events. 

As can be seen in Fig.~\ref{Fig:mGmG}, the kinematical reconstruction
of the LSP mass is possible if $m_X$ is sufficiently large,
$m_X\gsim 0.15 \mstau$ for $N_{\stau}=1000$ and $m_X\gsim 0.25
\mstau$ for $N_{\stau}=100$.  Otherwise one can get only an upper
bound on the mass $m_X$.

\subsection{measurement of the "Planck scale" }
\label{subsec:Mpl}

\begin{figure}[t!]
\begin{center}
\includegraphics{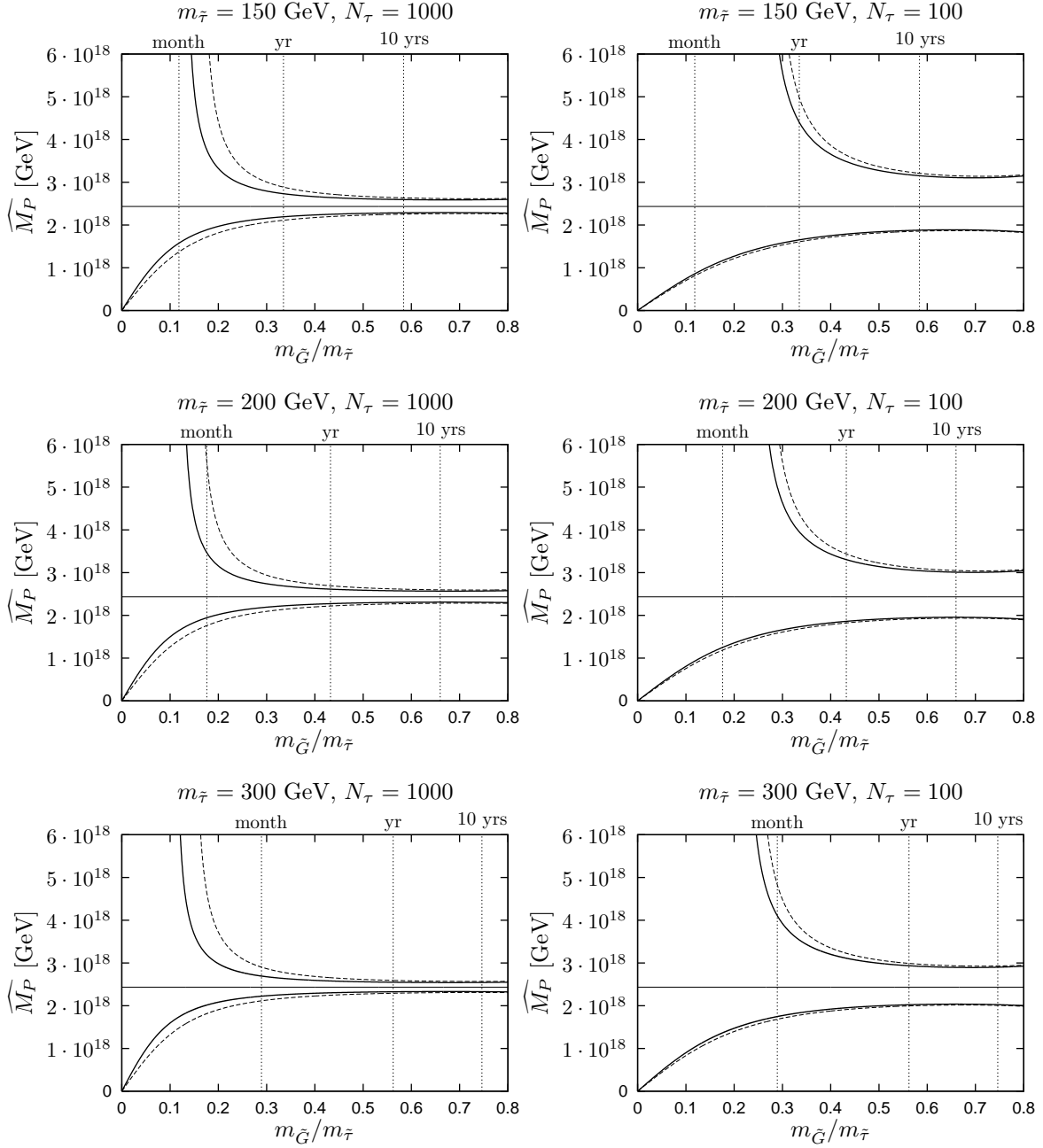}
\end{center}
\caption{Uncertainty of the Planck scale measurement, corresponding to
the estimated $1\sigma$ uncertainty of $E_\tau$, for $N_\tau = 1000$
and 100, and $\mstau = 150$, 200, and 300~GeV.  Solid lines represent
the case without an error of $\mstau$ measurement, and dashed lines
include 1\% uncertainty of $\mstau$. Vertical dotted lines represent
the stau lifetime of 1 month, 1 year, and 10 years.}
\label{Fig:mGMpl}
\end{figure}

Finally, if the LSP is the gravitino ($X=\gravitino$), the uncertainty
of the reconstructed gravitino mass $\widehat{\mgravitino}$ translates
into an uncertainty of the supergravity Planck scale, which is
obtained by substituting $\widehat{\mgravitino}$ for $\mgravitino$ in
Eq.(\ref{eq:Mp}). To take into account the error of $\mstau$, we also
substitute $\widehat{\mstau}$ for $\mstau$. Eq.(\ref{eq:Mp}) then
becomes
\begin{eqnarray}
  \widehat{\Mpl}^2 &=&
  \frac{1}{3\pi \Gamma_{\stau}}
    \frac{
      \left(\widehat{\mstau}\widehat{E_\tau} - m_\tau^2\right)
      \left(\widehat{E_\tau}^2 - m_\tau^2\right)^{3/2} 
    }{
      \widehat{\mstau}^2 + m_\tau^2 - 2 \widehat{\mstau} \widehat{E_\tau}
    }
    \:,
\label{calmpl}
\end{eqnarray}
which is shown in Fig.~\ref{Fig:mGMpl} for $\widehat{\mstau}=\mstau$
(solid lines) and $0.99 \mstau\le \widehat{\mstau}\le 1.01\mstau$
(dashed lines). Here we do not include the uncertainty of the lifetime
measurement, which simply affects the measured Planck scale as
$\Mpl\propto (1/\Gamma_{\stau})^{1/2}$. Neglecting the $\tau$--lepton
mass, Eq.~(\ref{calmpl}) is simplified as
\begin{eqnarray}
  \widehat{\Mpl}^2 &\simeq&
  \frac{1}{3\pi \Gamma_{\stau}}
    \frac{ \widehat{E_\tau}^4 }{ \widehat{\mstau} - 2 \widehat{E_\tau} }    
    \:.
\end{eqnarray}

As discussed in the previous section, the kinematical reconstruction
of the gravitino mass is possible only if $\mgravitino$ is
sufficiently large,
$\mgravitino\gsim 0.15 \mstau$ for $N_{\stau}=1000$ and $\mgravitino\gsim
0.25 \mstau$ for $N_{\stau}=100$.  For smaller values of the gravitino
mass, one can get only a lower bound on the Planck scale $\Mpl$.

One can see whether the determined `Planck scale' is inconsistent with
the Planck scale of the Einstein gravity, $\Mpl = 2.4 \times
10^{18}$~GeV. In other words, one can test the assumption of the decay
$\stau\to\tau\gravitino$ by comparing the observed lifetime with the
predicted lifetime.  For example, if the NLSP dominantly decays into
axino, the lifetime would be far shorter compared with gravitino
assumption for a fixed mass $m_X$, leading to a smaller value of
fitted $\Mpl$. Suppose for instance one measures $\mstau=150$~GeV and
$m_X=30$~GeV. When $X = \axino$, the lifetime becomes
$\mathcal{O}(10~\mathrm{sec})$ for $f_a\simeq 10^{11}~\mathrm{GeV}$
and $m_{\tilde{B}}\simeq \mstau$ [cf. Eq.~(\ref{eq:axinoGamma})]. If
one uses Eq. (\ref{calmpl}), a ``Planck scale''
$\widehat{\Mpl}=\mathcal{O}(10^{15}~\mathrm{GeV})$ would be obtained, 
thereby falsifying the gravitino assumption.

\subsection{model points and cosmological constraints}

We now discuss mSUGRA model points discussed in section
\ref{sec:modelpoints}, together with cosmological constraints.  In the
early universe, the $\stau$ CNLSP has been in thermal equilibrium
until its decoupling, $T_d\sim 0.04 \mstau$. If such a particle has a
very long lifetime, as discussed in this paper, its decay during or
after the big-bang nucleosynthesis (BBN), $T_\mathrm{BBN}\sim 1$~MeV,
may spoil the successful prediction of
BBN~\cite{BBNearly,Kawasaki:2004qu+Jedamzik:2006xz}. In the models with $\stau$ NLSP
and $\gravitino$ LSP, this leads to severe constraints on the
parameter space of ($\mstau$, $\mgravitino$), in particular to upper bounds on the gravitino mass
for a given stau mass~\cite{Asaka:2000zh+Fujii:2003nr,Feng:2004mt+Steffen:2006hw}.

Furthermore, it has been recently pointed out that a heavy charged particle can form a bound state with light elements during the BBN, which can lead to new effects and/or severer constraints~\cite{Pospelov:2006sc+Kohri:2006cn+Kaplinghat:2006qr}. In this paper we do not discuss these effects because it is difficult to evaluate the net effect quantitatively and it still awaits detailed analysis.

We should also mention that those BBN constraints may
disappear if there is entropy production between the stau
decoupling ($T_d\sim \mstau/20$) and the BBN ($T_\mathrm{BBN}\sim
1$~MeV), because the stau abundance is diluted before 
its decay~\cite{Buchmuller:2006tt}.

Keeping in mind the possibilities of severer bounds and also a possible loophole,
let us discuss the cases of $\mstau = 150$~GeV and $\mstau = 340$~GeV, corresponding to
the mSUGRA model points $\epsilon$ and $\zeta$. According to the latest analyses~\cite{Feng:2004mt+Steffen:2006hw} including the effects of the hadronic decay~\cite{Kawasaki:2004qu+Jedamzik:2006xz}, the bounds on the gravitino are
$\mgravitino\lsim (20$--$80)~\mathrm{GeV}$ for $\mstau\simeq 150~\mathrm{GeV}$ and 
$\mgravitino\lsim (40$--$200)~\mathrm{GeV}$ for $\mstau\simeq 340~\mathrm{GeV}$.\footnote{These constraints 
were derived without using the bound on the $^3$He. If one adopts it, the constraints become severer 
(cf.~\cite{Kawasaki:2004qu+Jedamzik:2006xz}).} 
The ranges of upper bounds correspond to the uncertainties of various bounds from primordial light elements. 

For $\mstau=150$ GeV (model point $\epsilon$),
as can be seen in Table~\ref{masssugra}, one could collect more than 1000 staus for 300 fb$^{-1}$.
The bound $\mgravitino\lsim
(20$--$80)$~GeV then suggests that the measurement of the gravitino mass
and the ``Planck scale'' may become possible if one assumes conservative BBN bounds
and if the gravitino mass is sufficiently large (cf. Figs.~\ref{Fig:mGmG} and \ref{Fig:mGMpl}).
For $\mstau=340$~GeV (model point $\zeta$),
the measurement would become easier if one could collect the same number of staus
(cf. Figs.~\ref{Fig:mGmG} and \ref{Fig:mGMpl}).
However, from Table~\ref{masssugra} we find that expected number of stopped CNLSP is around 30 for 300 fb$^{-1}$.
This is because gluino mass is above 2 TeV for this point and production cross section is
small. One needs SLHC ($\int {\cal L}=1000$fb$^{-1}$) to collect
$\mathcal{O}(100)$ events.

In the case of axino LSP, the BBN bound is much weaker because the
lifetime of the CNLSP stau becomes much shorter [see
Eq.(\ref{eq:axinoGamma})]. Hence, the axino mass measurement is
plausible for sufficiently large $\maxino/\mstau$.

\section{Light Axino vs gravitino: The rare decay of the CNLSP}
\label{sec:three-body}

\subsection{Low energy effective action of the axino and gravitino involving photon}
When $\mgravitino\lsim 0.2 \mstau$, 
it is difficult to determine
the gravitino mass from  $\tau$ energy measurement at 
stopper--detector. 
Axino is a motivated candidate  which couples weakly 
to the MSSM particle with comparable strength to the gravitino. 
In this section  we therefore  
compare the  decay $\stau\rightarrow 
\axino \tau\gamma$ with  the decay 
$\stau\rightarrow \gravitino\tau \gamma$.
For simplicity, we will assume the NLSP is pure `right-handed' stau,
$\stau = \tilde{\tau}_R$ throughout this section. Extension to the
case with mixing with $\tilde{\tau}_L$ is straightforward, 
and  the mixing angle dependence
is expected to be small.

The  gravitino is a spin-3/2 particle. However,  in the limit where 
$\mgravitino \ll \mstau$,  
the effective interaction to  MSSM particles would be reduced 
to that of spin 1/2  particle, goldstino $\tilde{\chi}$. The 
effective action  relevant to the $\tilde{\tau}_R$ decay is given as follows,
\begin{equation}
L=
\frac{\mstau^2}{\sqrt{3}\mgravitino\Mpl}
\left(\bar{\tilde{\chi}}\tau_R\tilde{\tau}^*_R
+ H.c.\right)
+\frac{-m_{\tilde{B}}}{4\sqrt{6}m_{3/2}\Mpl}\bar{\tilde{\chi}}
[\gamma^{\mu},\gamma^{\nu}]\tilde{B} 
(\cos\theta_WF_{\mu\nu}
-\sin\theta_WZ_{\mu\nu}). 
\label{eq:actiongv}
\end{equation}
The action is similar to that of axino given in
Eqs.(\ref{eq:AxinoBino}) and (\ref{eq:AxinoStau}) except the coupling
coefficients.  The relative weight of the two terms in
Eq.(\ref{eq:actiongv}) are fixed by the supergravity, while for the axino
the coefficient in Eq.(\ref{eq:AxinoStau}) is induced from
Eq.(\ref{eq:AxinoBino}) by the radiative corrections.  Note that the
term proportional to $X[\gamma^{\mu}, \gamma^{\nu}]\tilde{B}F_{\mu\nu}$
is a non renormalizable coupling of the photon to gravitino or axino and
induces significantly different $\gamma$, $\tau$ distribution.

\begin{figure}[t]
\begin{center}
\includegraphics[width=10cm]{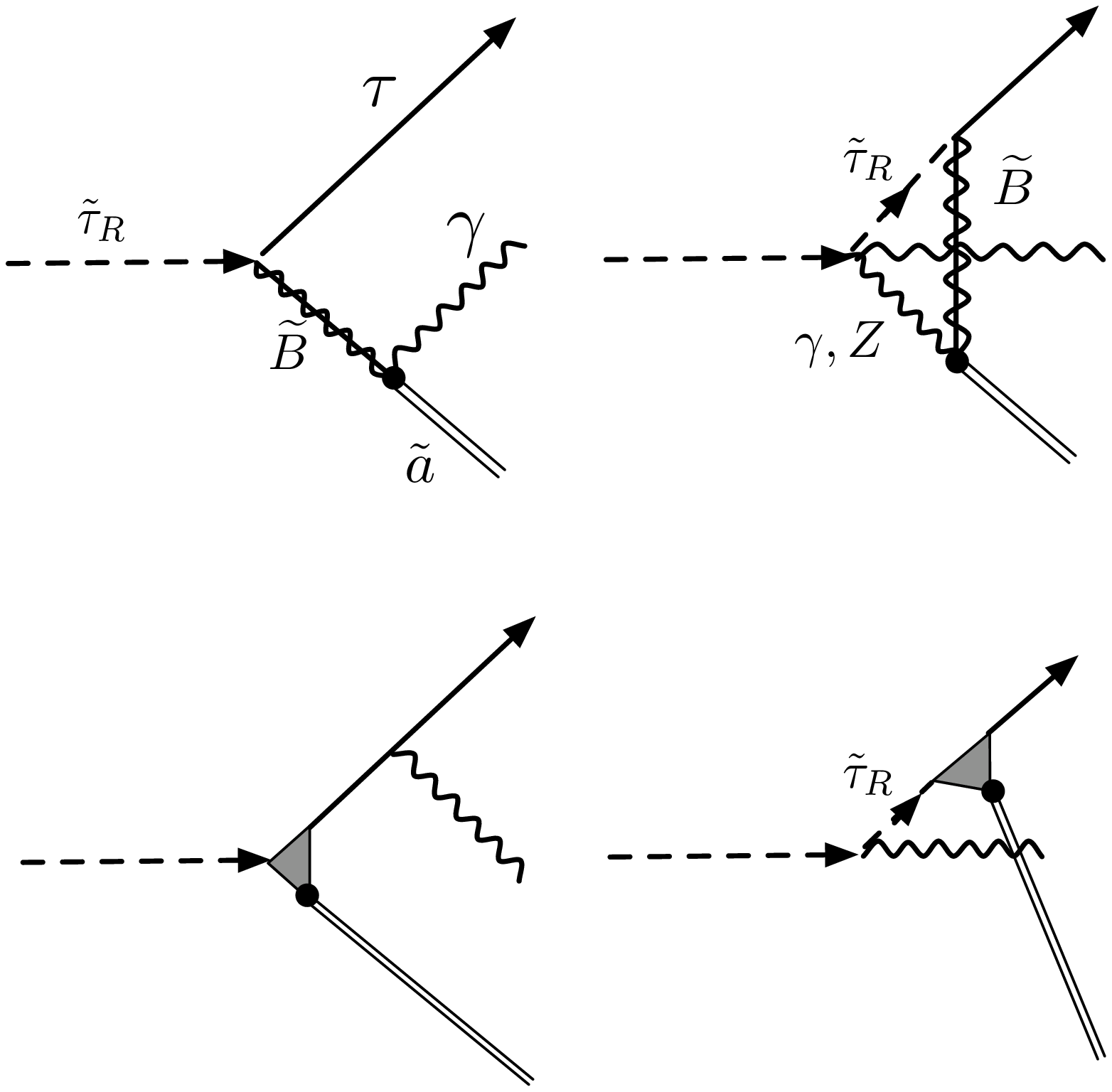}
\end{center}
\caption{Leading Feynman diagrams for $\stau\rightarrow \tau \gamma \axino$. 
}
\label{Fig:threebody}
\end{figure}

\begin{figure}[t]
\begin{center}
\includegraphics[width=10cm]{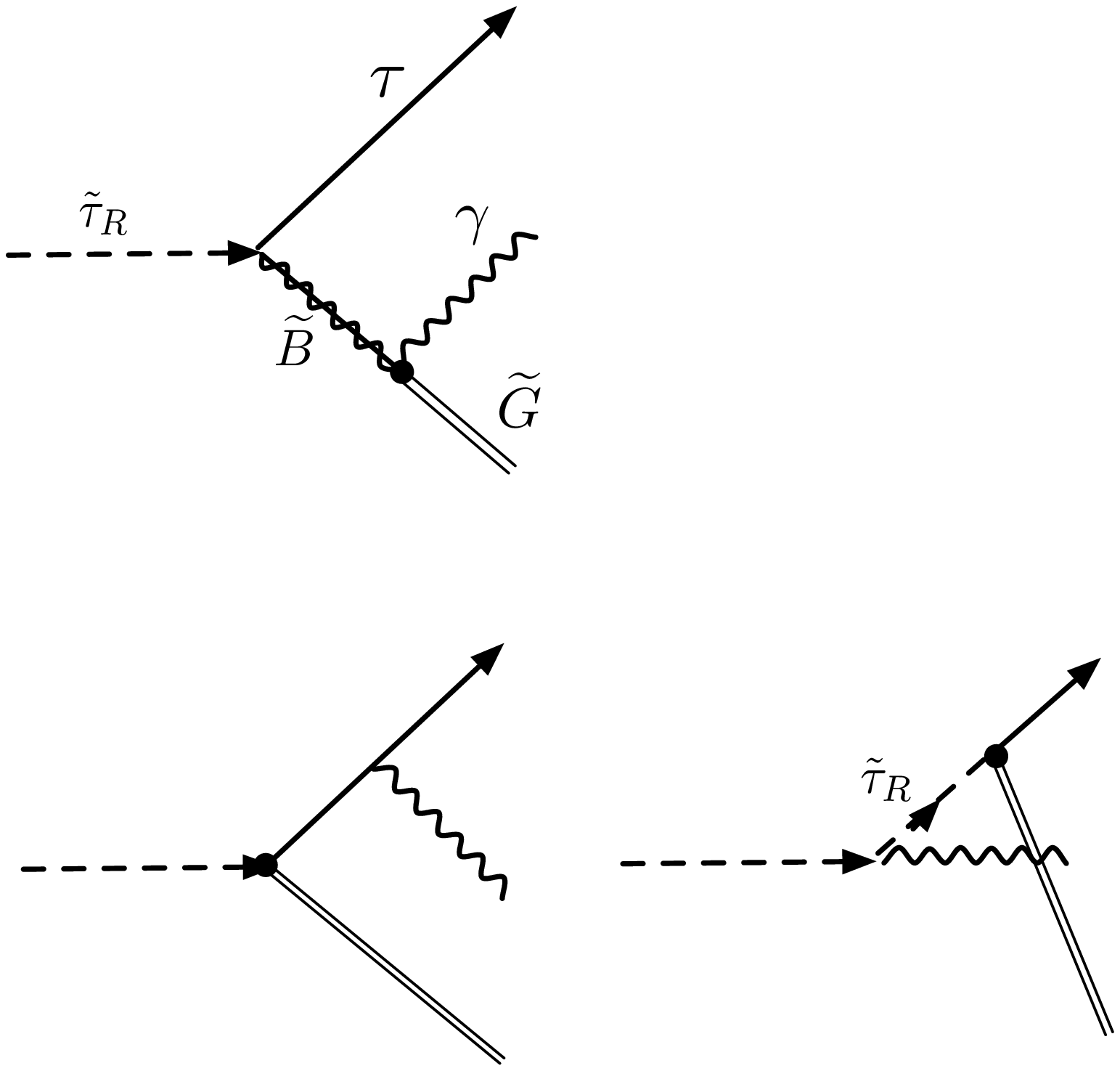}
\end{center}
\caption{ Feynman diagrams for $\stau\rightarrow \tau
\gamma\gravitino$ in the goldstino limit. 
}
\label{Fig:threebody_gra}
\end{figure}

The axino three body decay $\stau\rightarrow \gamma\tau\axino$
proceeds through the diagrams shown in Fig.\ref{Fig:threebody}, where the
hatched triangle express the effective vertex shown in
Eq. (\ref{eq:AxinoStau}). On the other hand, 
the
relevant diagrams for the three body decay  into goldstino  $\stau\rightarrow\gamma\tau
\tilde{\chi}$, are given in Fig.\ref{Fig:threebody_gra}.  
The diagram
corresponding to the top right of Fig. \ref{Fig:threebody} does 
not exist for the goldstino case.  
The difference of the actions and the relevant diagrams 
will appear  as the deviation of the decay distributions. In the
Appendix, we list the three body decay differential width into
gravitino/axino in the limit where the gravitino/axino mass can be
neglected compared to $\mstau$. The formula for the massive
gravitino and axino are given in Ref.~\cite{Brandenburg:2005he}.

\subsection{Numerical results} 
  
The three body decay $\stau\to \tau\gamma X$ should be visible in the
stopper--detector if it has an ability to measure charged tracks, and
also segmented into small units. The position where $\stau$ decays in
the detector is the position where the charged track by the $\pi^+$,
$\mu$ and $e$ from $\tau$ is initiated.  For hadronic tau decays, a
$\pi^{\pm}$ is always in the decay products, sometimes with photons
coming from $\pi^0$ decays.  Photons are converted into electron after
passing $\sim $ one radiation length $X_0$.  Therefore for the iron
based detector with $\rho=5(2)$g/cm$^3$, the photon shower starts
$2.8(7)$cm  from the decay point.  In summary, the
three body decay of $\stau\rightarrow\tau\gamma X$ is identified as a
charged track (which might be associated with collinear EM showers) +
an isolated hard EM shower pointing back to the point where charged
track is started.  If the segmentation is not good enough, the
efficiency to discriminate EM showers from $\pi^{\pm}$ would be
reduced.

As can be seen in the appendix, the three body decay amplitude can be
written as a function of the angle between photon and $\tau$,
$\theta$, and $E_{\gamma}$. To be conservative we assume the energy resolution
for the isolated photon shower is 
$\Delta E_{\gamma}/E_{\gamma}=100\%/\sqrt{ E_{\gamma}/{\rm GeV } }$ 
and ignore the
angular resolution of the photon momentum. (In the following
$E_\gamma$ denotes the photon energy after taking into account of 
this finite resolution effect.) Note that the shortest length of the
detector is 3.5 m, which corresponds to 1750g/cm$^2$ 
for $\rho=5$g/cm$^{3}$. The EM showers are likely contained in the
stopper because we only need 200g/cm$^2$ to fully absorb them.
 
  We need to require several cuts for the accepted events.  
\begin{itemize}
\item 
Experimentally, the angle between $\tau$ and $\gamma$ must be
large enough to avoid the overlap between 
$\tau$ decay products and prompt $\gamma$.  We only use the 
events where $\cos\theta < 0.866$. 
We do not lose sensitivity to the differece between 
the two decays $\stau\rightarrow \tau \gamma \axino/\gravitino$ 
by cutting these events. 
In the collinear region, the amplitude is 
dominated by the contributions from  QED vertex, 
which is common for both  of the decays. 

\item 
The three body decay amplitude suffers soft and
collinear singularity. Because we only adopt simple leading order
calculation, we require $E_{\gamma}>$ 10~GeV and 
$E_\tau>10$~GeV. 
\end{itemize}

We define the ratio 
\begin{equation}
R(X)=
\Gamma(\stau\rightarrow 
X\gamma\tau)\vert_{\rm after \ cut}/
\Gamma(\stau\rightarrow X\tau)
\end{equation}
The dependence of $R$ on the MSSM parameter 
is quite different between $X=\axino$ and $X=\gravitino$.
While $\stau\rightarrow \axino \tau $ 
is one loop process controlled by the parameter $\xi$,
 the  three body decay contains 
a  tree level contribution which depends on non-renormalizable 
axino-$\tilde{B}$-gauge coupling (top left of  Fig. \ref{Fig:threebody}).

\begin{figure}[t]
\begin{center}
\includegraphics[width=7cm, angle=90]{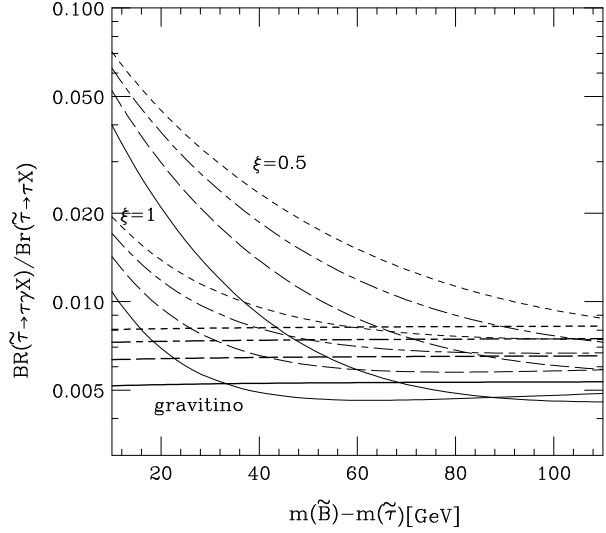}
\end{center}
\caption{
The ratio $R(X)=\Gamma(\stau\rightarrow X\gamma 
\tau)/\Gamma(\stau\rightarrow X \tau)$  as 
a function of $m_{\tilde{B}}-\mstau$. 
The solid, dashed, dot-dashed, short-dashed curves correspond to 
the case where $\mstau=100, 130, 160, 190 $~GeV
respectively.  Lines which increase toward smaller mass 
difference corresponds to $X=\axino$. The $X=\gravitino$ 
lines do not show significant bino mass dependence.
}
\label{Fig:br}
\end{figure}

When $\xi$ is small, the tree level contribution  plays a dominant 
role in the three body decay into axino. 
This can be seen in Fig. \ref{Fig:br}, where the ratio $R(\axino)$ 
is plotted as a function of the mass difference 
$m_{\tilde{B}}-\mstau$. We also fix $m_{X}=1$~GeV but 
$R(X)$ is insensitive to $m_{X}$. 
$R(\axino)$ is enhanced  when the mass difference between 
$m_{\tilde{B}}$  and $\mstau$ 
is small relative to the $\stau$ mass. 
This is because 
the two  body decay of axino is suppressed 
by $m_{\tilde{B}}^2$. 
The branching 
ratio ranges from  8\% to 0.5\% for the model parameters 
given in the figure.   
On the other hand, 
the branching ratio $Br(\stau\rightarrow \tau
\gamma \gravitino)$ is well below 1\% for the parameter given in
Fig.\ref{Fig:br}.  $Br(\stau\rightarrow \tau \gamma
\gravitino)/Br(\stau\rightarrow
\tau\gravitino)\sim 0.56\%$ $ ( 0.84\%)$ for $\mstau=100(190)$~GeV 
respectively. 

\begin{figure}[t]
\begin{center}
\includegraphics[width=7cm, angle=90]{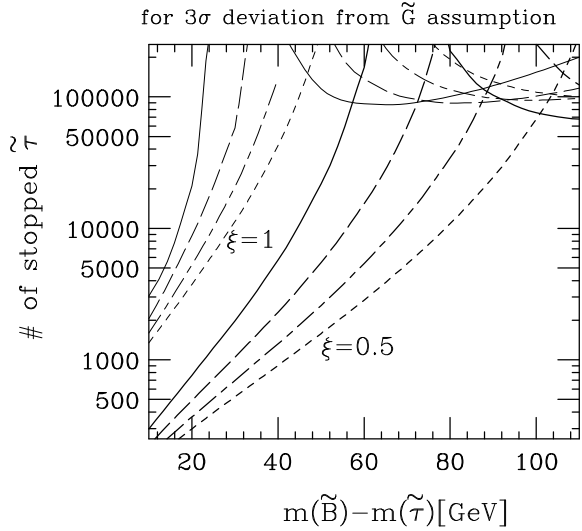}
\end{center}
\caption{ The statistics required to find the branching ratio 
$R(\axino)=\Gamma(\stau\rightarrow \tau \gamma\axino)
/\Gamma(\stau\rightarrow \tau \axino)$  deviates more than 3 sigma 
from the prediction for $R(\gravitino)$.    
The lines corresponds to different stau mass as in Fig.\ref{Fig:br}. }
\label{Fig:total}
\end{figure}

When  the three body decay branching ratio turns out 
be above 5\%, 200 stopped $\stau$ is enough to see the 
 3 $\sigma$ deviation from the 
 gravitino assumption. We estimate the number of 
events $N_{\rm  event}$ stopped in the detector, which is 
required to find  3$\sigma$ deviation from the gravitino 
scenario as follows; 
\begin{equation}
N_{\rm event}\frac{(Br( \stau\rightarrow \tau\gamma\axino)-Br
( \stau\rightarrow \tau\gamma\gravitino))^2}{Br
(\stau\rightarrow \tau\gamma\axino)}=9. 
\end{equation}  
The $N_{\rm event}$ as a function of $m_{\tilde{B}}-\mstau$ is given
in Fig.\ref{Fig:total}.  Each curve increases as
$m_{\tilde{B}}-\mstau$ is increased up to the value where
$Br(\tilde{\tau} \rightarrow \tau\gamma \axino)$ coincides with
$Br(\tilde{\tau}\rightarrow \tau \gamma\gravitino)$.  When
$m_{\tilde{B}}-\mstau<20$~GeV, $\mathcal{O}(1000)$ (
$\mathcal{O}(10000)$ ) stopped $\stau$ are enough to see the deviation
from the measurement of $R$ for $\xi=0.5(1)$ respectively.

\begin{figure}[t]
\begin{center}
\includegraphics[width=5cm, angle=90]{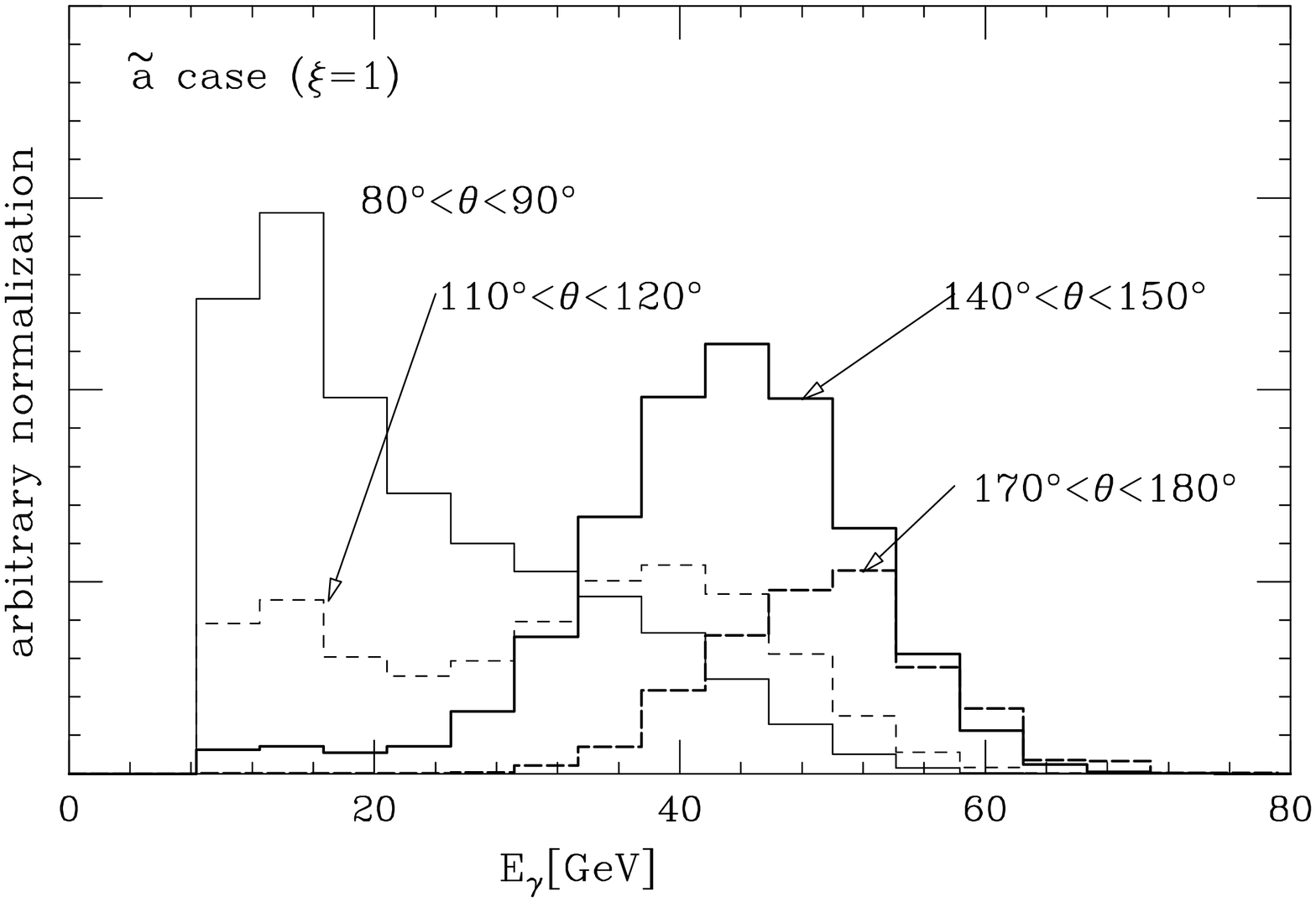}
\includegraphics[width=5cm, angle=90]{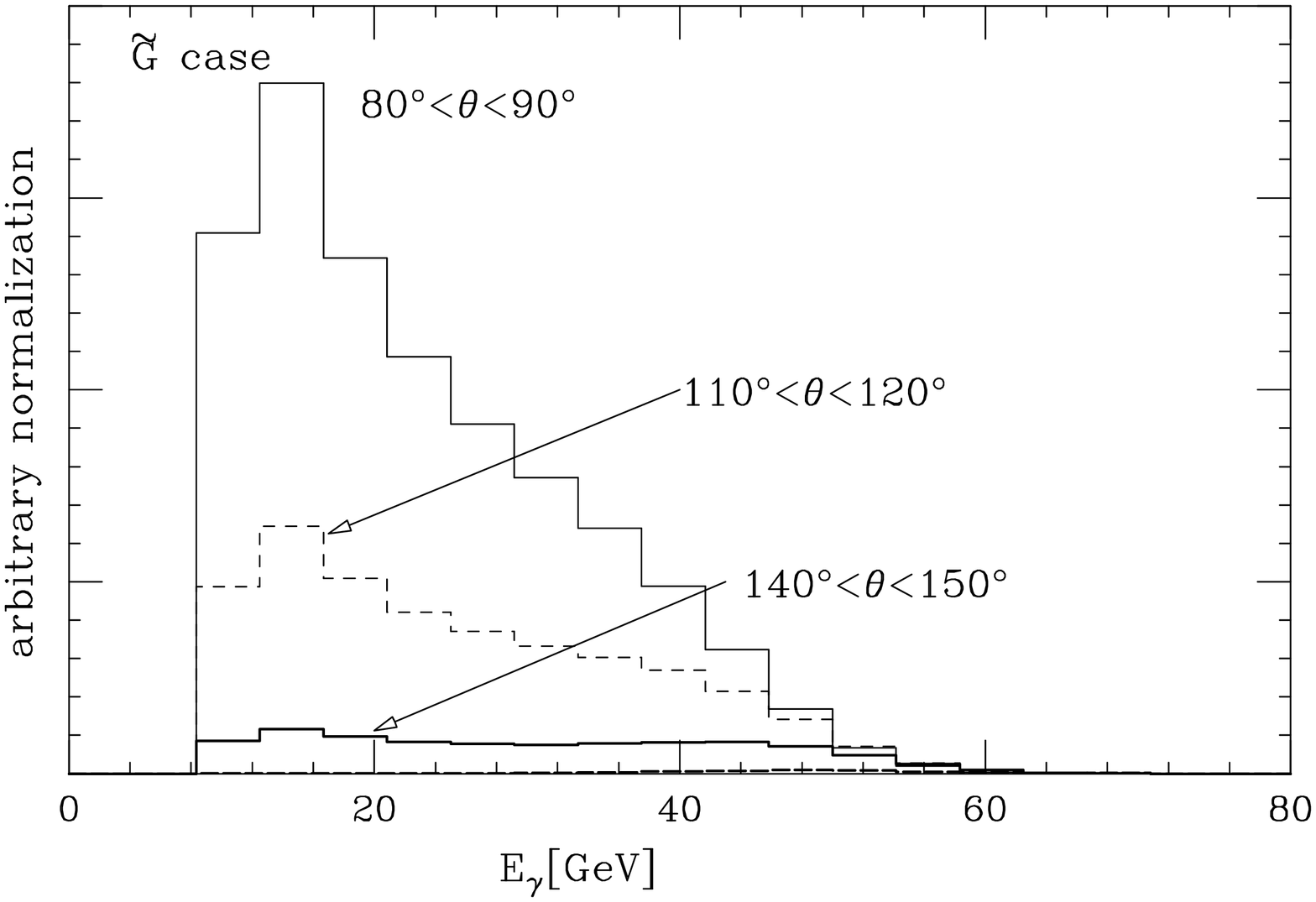}
\end{center}
\caption{The distribution of  $E_{\gamma} $ in  some $\theta$ intervals for 
$\stau\rightarrow\tau \gamma X$ decay.  left panel: $X=\axino $ $(\xi=1) $,  right panel: $X=\gravitino$. $\mstau=100$~GeV, and $m_{\tilde{B}}=130$~GeV. }
\label{Fig:egamdist}
\end{figure}

\begin{figure}[t]
\begin{center}
\includegraphics[width=6.8cm, angle=90]{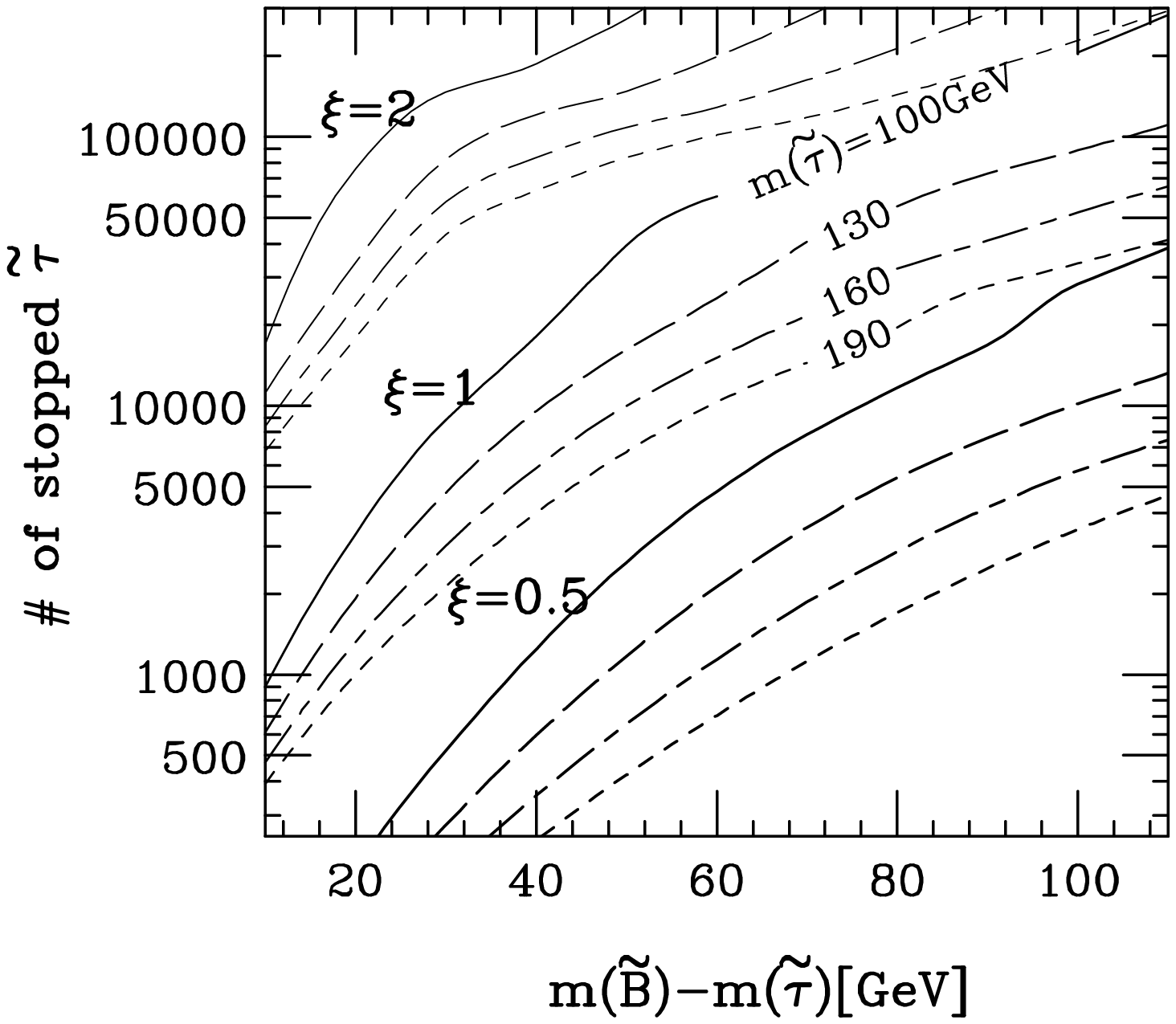}
\includegraphics[width=7cm, angle=90]{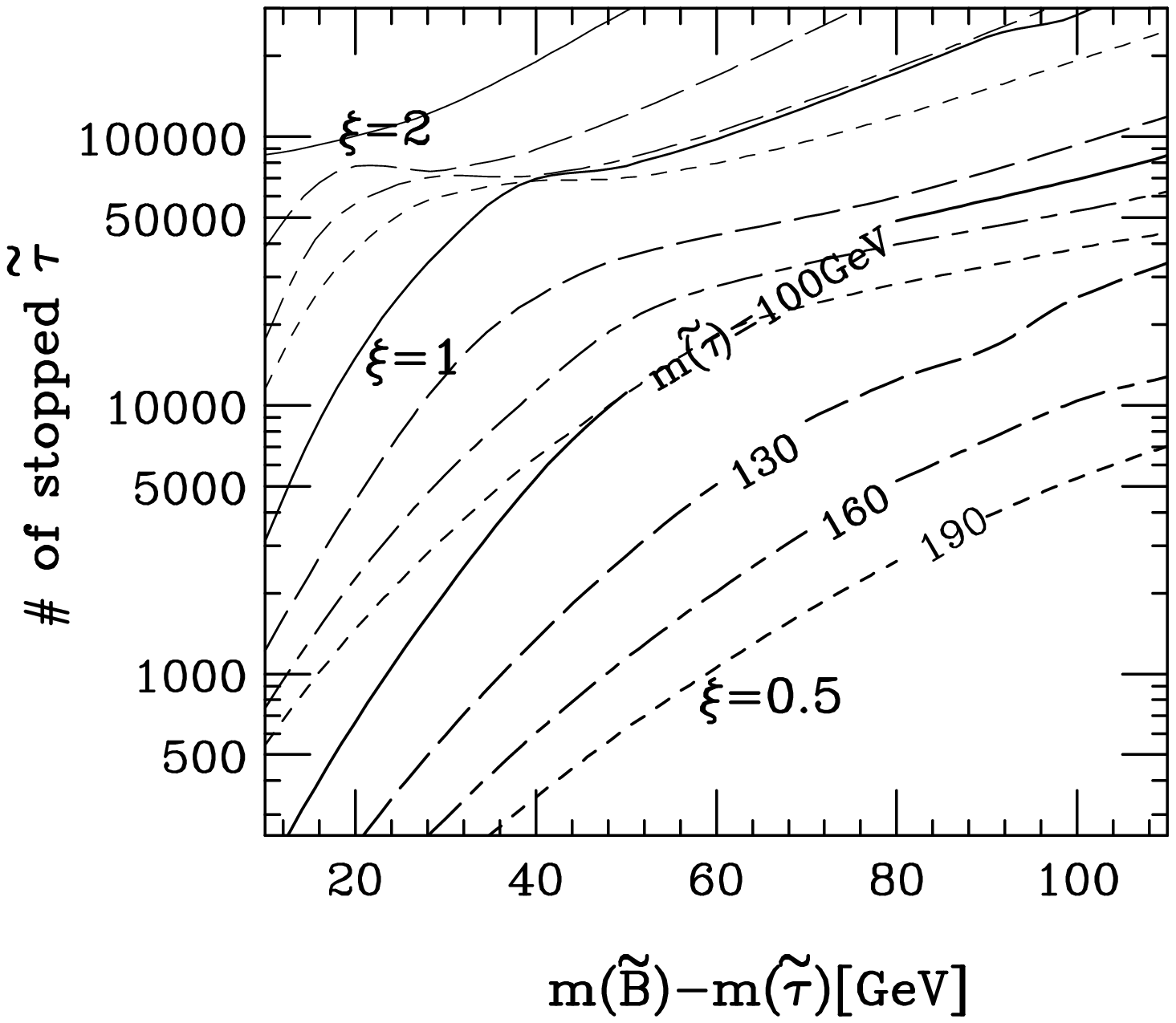}
\end{center}
\caption{
Statistics required to distinguish 
axino scenario from gravitino scenario. 
left for $\mgravitino =\maxino=1$~GeV
and right for $\mgravitino =\maxino=30$~GeV
}
\label{Fig:dist}
\end{figure}

Not only the branching ratio, but also the decay distribution
contain the information of the invisible particle. 
The axino decay distribution can be enhanced at the region where 
$x_{\gamma}=2E_{\gamma}/\mstau$  is large and 
$\cos\theta$ is small, namely 
hard photon and $\tau$ is back to back 
~\cite{Brandenburg:2005he}. 
This occurs when $\xi$ is small  and relative importance 
of the  direct axino-bino-gauge boson coupling is enhanced. 

In Fig. \ref{Fig:egamdist},  left panel   shows the 
$E_{\gamma}$  distribution for
different 
$\cos\theta$ intervals. 
Here we fix $\xi=1$, 
$m_{\tilde{B}}=130$~ GeV, and $\mstau=100$~GeV. 
The distribution 
has significantly hard component for 
$\cos\theta<0$ when compared with 
 that of gravitino  in the right panel.  The  enhancement 
of back-to-back events is  a  signature of an axino, 
and  clear $\tau$ and $\gamma$ 
separation  is not required to distinguish them.  
 
We  now estimate the the number of  CNLSP decays 
in the stopper  to see the 3 $\sigma$ deviation between 
$\stau\rightarrow \axino\gamma\tau $  and 
$\gravitino\gamma\tau$ from the decay distribution. We define 
$\Delta\chi^2$ like function from the differential
decay width;
\begin{equation}
\Delta\chi^2_{\rm dist} (\mstau, m_X, m_{\tilde{B}}, \xi, 
\Delta E_{\gamma}, \Delta \theta)
=\Sigma_i\frac{ 
(n_i(\axino)-\bar{n}_i(\gravitino))^2}{n_i(\axino)}. 
\end{equation}   
Here,  $n_i(\axino)$ is the number of event in a $i$-th bin 
for $\stau\rightarrow \tau \gamma \axino$ when 
the number of stopped  $\stau$ is   $N_{\rm gen}=10^5$. 
$\Delta E_{\gamma}$ and $\Delta \theta$ is the bin size in  
$E_{\gamma}$ and $\theta$. 
We divide the $E_{\gamma}$ and 
$\theta$ into 8 bins and 3 bins respectively, 
for the range  $0 <E_{\gamma}<\mstau$ and
$0<\theta<\pi$;
\begin{equation}
\Delta E = \mstau/8, \ \ \Delta \theta =\pi/3.
\end{equation} 
The photon energy resolution of the detector 
is assumed $\Delta E_{\gamma}/E_{\gamma}=100\%/\sqrt{E_{\gamma}/
{\rm GeV }}$.
We apply the cut  $\cos\theta<0.866$, $E_{\gamma}>10$~GeV 
and $E_{\tau}>10$~GeV.
On the other hand,  $\bar{n}_i(\gravitino)$ is the number of event in a $i$-th bin for 
$\stau\rightarrow \tau \gamma \gravitino$ with $\mgravitino = \maxino$ 
normalized so that  the total number is same to that of 
axino three  body decay. Namely, we do not use the information for 
$R(x)$ in our fit. 

From $\Delta\chi^2_{\rm dist}$,  
 we  define $N_{\rm dist}(3\sigma)$,  
the  number of stopped $\stau$  required to see 
the 3$\sigma$ deviation between  
$\stau\rightarrow \tau \gamma \axino$ and 
$\stau\rightarrow \tau \gamma \gravitino$ 
as follows, 
\begin{equation}
N_{\rm dist}(3\sigma) =N_{\rm gen}/(\Delta\chi_{\rm dist}^2/9). 
\end{equation}
We show the $N_{\rm dist}(\sigma)$ as a function of 
$m_{\tilde{B}}-\mstau$  in Fig. \ref{Fig:dist}. 
The sensitivity is significantly increased from the estimate using 
the branching ratio only. 
 For $m_{\tilde{B}}-\mstau=40$~GeV, 
the deviation may be visible for $\mathcal{O}(1000)$ events for $\xi=0.5$ 
($\mathcal{O}(10000)$ events for $\xi=1$).  

Finally, we estimate sensitivity at our model points 
in section 3. For simplicity we assume $\stau=\stau_R$.
In general 
$\tilde{\tau}_R$ is mixed with $\tilde{\tau}_L$. The mixing
angle is defined as \begin{equation}
\tilde{\tau}_1=\tilde{\tau}_L\cos\theta_{\tilde{\tau}}
+ \tilde{\tau}_R \sin\theta_{\tilde{\tau}}\;.
\end{equation}
For model points discussed in Table \ref{mass}, the angle is 
$\sin\theta_{\tilde{\tau}}\sim 0.9$.  The effect of the mixing angle
in the axino decay is small because the amplitude
of $\tau_L$ is suppressed by both by the small
$\cos^2\theta_{\tilde{\tau}}$ factor and  smaller 
hypercharge, and can be safely ignored.

\begin{figure}[t]
\begin{center}
\includegraphics[width=7cm, angle=90]{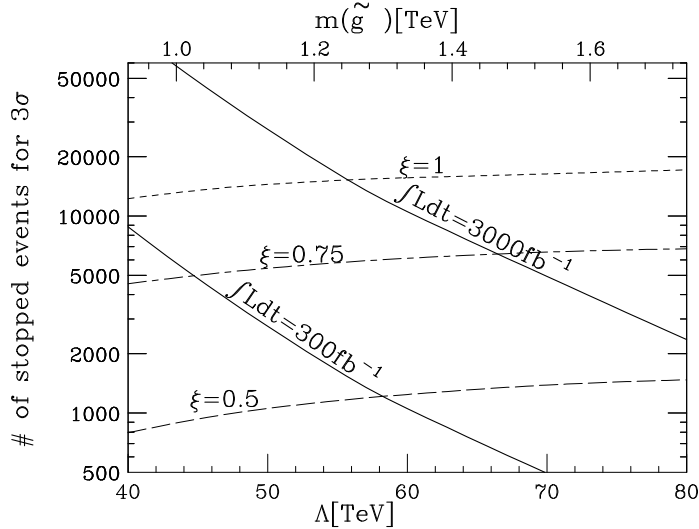}
\end{center}
\caption{Solid lines are the expected number of stopped events for 300(3000)fb$^{-1}$
luminosity for GM points with $40{\rm TeV}<\Lambda<80$TeV. Dashed lines 
are number of required stopped events to see 
3-$\sigma$ deviation in the $E_{\gamma}-\cos\theta$ distribution of 
 $\stau\rightarrow\tau\gamma\axino$  from those 
expected for the decay $\stau\rightarrow\tau\gamma\gravitino$.              
}
\label{Fig:lambda}
\end{figure}

In Fig.\ref{Fig:lambda} 
the expected sensitivity at the stopper--detector is shown. Here,  long-dashed 
(long-short-dashed, dashed) line corresponds to the required statistics 
for  $\xi=0.5$ (0.75, 1)  for different  $\Lambda$, 
while the upper and lower solid lines correspond
to the number of stopped CNLSP for 300 fb$^{-1}$ and 3000 fb$^{-1}$. 
One can address the difference between axino and gravitino
for $\Lambda\sim 55 (65)$TeV or less for   
${\cal L}=3000$ fb$^{-1}$ and $\xi=1$ (0.75).
Note that the expected reach in $\Lambda$   is essentially determined 
by the SUSY production cross section, as they decrease  steeply 
with increasing $\Lambda$.  We therefore 
show the gluino masses scale on the top of the figure.

\section{Discussion} 
\label{sec:discussion}

In this paper we investigate the physics of the long-lived charged
next lightest SUSY particle (CNLSP), which may be explored at a
massive stopper--detector placed next to the CMS detector at the LHC.
We assume the CNLSP is the lighter scalar tau, $\tilde{\tau}$, which
decay into $\tau X $ where $X$ is a invisible particle.  A natural
candidate of the particle $X$ is a gravitino but we also consider the
case where $X$ is axino $\axino$.  

In this paper,  we assume  very large stopper--detector next to the 
CMS detector, with total mass 
of 8 kton.   The stopper must have a capability to measure the position where 
the  CNLSP stopped, and also the 
energy of the $\tau$ decay products. If the detector can be  highly segmented, 
it is also  possible to identify  the $\tau$ decay products separately. 
The number of stopped  NLSP ranges from $\mathcal{O}(10^4)$ 
to $\mathcal{O}(100)$ for gluino and squark with mass below  2 TeV. 
If the size of the detector should be smaller,  the number of the events 
must be scaled down linearly. 

We estimate the statistical 
error for $E_{\tau}$, which can be determined from the end point of 
the tau jet energy.  We assume that 
the energy resolution of the detector is 
150\%/$\sqrt{E/{\rm GeV}}$.  
$\delta E_{\tau}/E_{\tau} \simeq 60\%/\sqrt{E_\tau/{\rm GeV}}$
($15\%/\sqrt{E_\tau/{\rm GeV}}$)
can be achieved for the statistics of the order 
of $N_{\rm stopped}=100(1000)$.  
 When LSP  mass is above $0.2\mstau$, one can 
 constrain the gravitino mass both from above and below. 
From the lifetime and gravitino mass measurement, 
the supergravity Planck scale  $\Mpl $ 
can be measured.
If many CNLSP can be accumulated, one can even study the three body 
decay of the CNLP $\tilde{\tau}\rightarrow \tau\gamma X$. 
When the dominant decay mode of the 
 $\tilde{\tau}$ is $\tilde{\tau}\rightarrow \axino\tau$,  we may be able to 
see the deviation of the decay branching ratio and the distribution from 
the $X=\gravitino$ case. 

Finally we comment on the strategy to proceed this experiment.  SUSY
particles will be found in the early stage of the LHC experiment if
SUSY scale is $\mathcal{O}(1~\mathrm{TeV})$.  If LSP is gravitino or
axino and the NLSP is charged and long-lived, it would also be
recognized easily.  The detector proposed in this paper may be placed
after the existence of long-lived CNLSP is observed, roughly at the
same time to the high luminosity run of the LHC, or the proposed super
LHC run.

LHC experiment has a great potential to explore new physics in TeV
regions.  It is important to explore new possibilities that can be
done with LHC.  In this paper we have proposed a large additional
detector in the CMS cavern when the long-lived CNLSP is found.  This
requires a significant modification of the CMS experiment.  The reward
is low systematics study of the CNLSP decay which primarily serve for
the study of the gravitino sector.  The determination of gravitino
mass either from the lifetime or (independently) from $E_{\tau} $
measurement would give us a direct information of the the total SUSY
breaking scale.  Together with high precision determination of the MSSM
sector expected with the CNSLP momentum information, the nature of the
interaction of the MSSM sector and hidden sector can be studied in
detail.  We hope this paper is useful for further, and more realistic
studies.

\section*{Acknowledgement}
We thank J.~Ellis, H.~U.~Martyn and F.~D.~Steffen for discussions.
This work is supported in part by the Grant-in-Aid for Science Research,
Ministry of Education, Science and Culture, Japan (No.16081207, 18340060)
for MMN.

\section{appendix}  
In  this appendix we list the differential decay width of $\tilde{\tau}_R$ 
decay 
into $\gamma\tau X$ where $X$ is either gravitino or axino in limit 
where $m_X\ll m_{\tilde{\tau}_R}$. 
Formulas in this appendix are obtained by taking massless limits
$\mgravitino / \maxino \to 0$ of the formulas in
Ref.~\cite{Brandenburg:2005he}. The
$\tilde{\tau}_R$ decay width to $\tau\gamma \gravitino$ is given as follows;
\begin{equation}
\frac{d^2\Gamma(\tilde{\tau}_R\rightarrow \tau\gamma \gravitino)}
{dx_{\gamma} d\cos\theta}
=\frac{\mstau}{512\pi^3}
\frac{x_{\gamma}(1-x_{\gamma})}{[1-(x_{\gamma}/2)(1-\cos\theta)]^2}
\sum_{\rm spins} \vert {\cal M}(\tilde{\tau}_R\rightarrow \tau\gamma \gravitino)\vert^2 
\end{equation}
where 
\begin{equation}
\sum_{\rm spins}\vert M(\tilde{\tau}_R\rightarrow \tau \gamma
\gravitino)\vert^2
=\frac{8\pi\alpha}{3}\frac{m^2_{\tilde{\tau}}}{\Mpl^2 A_{\gravitino}}
F^{\gravitino}_{\rm diff} (x_{\gamma}, \cos\theta, A_{\tilde{B}}).
\end{equation}
and 
\begin{eqnarray}
F^{\gravitino}_{\rm diff}
&=&
\frac{1+\cos\theta}{1-\cos\theta}
\left[
-x_{\gamma}+\frac{2}{x^2_{\gamma}}+\frac{1}{1-x_{\gamma}}
\right.
\cr
&&
\left.
+1+\frac{4x_{\gamma} -2(1-\cos\theta) -4}{2-x_{\gamma}(1-\cos\theta)}
+\frac{4-4x_{\gamma}}{\{2-x_{\gamma}(1-\cos\theta)\}^2}
\right]
\cr
&&+\frac{1-\xg}{2-\xg(1-\ct)}
\left(
-4+2x-\frac{4}{\xg}
\right)
+
\frac{4(1-\xg)^2}{\{2-\xg(1-\ct)\}^2}
\cr
&&
+\frac{1}{\{\bunbo\}^2}\Biggl[
\cr
&&
~~
2\ab^2(1-\xg)
\left\{\xg^2-2\xg+\frac{4}{\xg}-(\xg^2-2\xg-2)\ct\right\}
\cr
&&
~~
+\ab(1+\ct)(4-3\xg+\xg\ct)(\xg^2-\xg-2) 
\cr
&&
~~
-(1+\ct)^2\xg(\xg^2-\xg-2)
\Biggr]
.
\end{eqnarray}
Here 
\begin{eqnarray}
x_{\gamma}=2E_{\gamma}/\mstau, \ \ A_{\tilde{B}}= (m_{\tilde{B}}/\mstau)^2, 
A_{\gravitino}= (\mgravitino/\mstau)^2, 
\end{eqnarray}
and $\theta$ is the angle between $\tau$ and $\gamma$.

For the case of massless axino, we find 
\begin{equation}
\frac{d^2 \Gamma(\tilde{\tau}_R \rightarrow \tau \gamma \axino)}
{dx_{\gamma}d\cos\theta}=\frac{\mstau}{512\pi^3}
\frac{x_{\gamma}(1-x_{\gamma})}{[1-(x_{\gamma}/2)(1-\cos\theta)]^2}
\sum_{\rm spins}\vert {\cal M}(\tilde{\tau}_R\rightarrow \tau\gamma\axino)\vert^2,
\end{equation} 
where 

\begin{equation}
\sum_{\rm spins}\vert M(\tilde{\tau}_R\rightarrow \tau\gamma
\axino)\vert^2
=
\frac{
\alpha^3
C^2_{\axino YY}}{\pi
\cos^4\theta_W
} 
\frac{\mstau^2}
{f^2_{a}} F^{\axino}_{\rm diff}
(x_{\gamma}, \cos\theta,A_{\tilde{B}}),
\end{equation}
\begin{eqnarray}
F^{\axino}_{\rm diff}
(x_{\gamma},\cos\theta,A_{\tilde{B}})
&=&\frac{\xg^2(1-\xg)(1+\ct)\{1+\ct+\ab(1-\ct)\}}{\{\bunbo\}^2}
\cr
&&+A\frac{\ab(1+\ct)}{\bunbo}
\cr
&&+\frac{1}{4}A^2\frac{1+\ct}{1-\ct}\ab
\left(\frac{1}{1-\xg}+\frac{2}{\xg^2}
\right),
\end{eqnarray}
\begin{equation}
A=\frac{3\alpha}{\pi\cos^2\theta_W}\xi\log\left(\frac{f_a}{m}\right).
\end{equation}

\end{document}